\documentclass[aip,jcp,reprint]{revtex4-1}

\usepackage{graphicx}
\usepackage{dcolumn}
\usepackage{bm}
\usepackage{amssymb}
\usepackage{rotating} 
\usepackage{amsmath}

\newcommand{\sub}[1]{\ensuremath{_{\textrm{#1}}}}
\newcommand{\one}{\:^1}
\newcommand{\two}{\:^2}
\DeclareMathOperator{\Tr}{Tr}
\DeclareMathOperator{\rank}{rank}

\begin{document}

\title{Global Solutions of Restricted Open-shell Hartree-Fock Theory
  from Semidefinite Programming with Applications to Strongly
  Correlated Quantum Systems}

\author{Srikant Veeraraghavan and David A. Mazziotti}

\affiliation{Department of Chemistry and The James Franck Institute, The University of Chicago, Chicago, IL 60637}

\email{damazz@uchicago.edu}

\date{Submitted December 20, 2013; Revised February 12, 2014}

\begin{abstract}

  We present a density matrix approach for computing global solutions of restricted open-shell Hartree-Fock (ROHF) theory, based on semidefinite programming (SDP), that gives upper and lower bounds on the Hartree-Fock energy of quantum systems.  While wave function approaches to Hartree-Fock theory yield an upper bound to the Hartree-Fock energy, we derive a semidefinite relaxation of Hartree-Fock theory that yields a rigorous lower bound on the Hartree-Fock energy.  We also develop an upper-bound algorithm in which Hartree-Fock theory is cast as an SDP with a nonconvex constraint on the rank of the matrix variable.  Equality of the upper- and lower-bound energies guarantees that the computed solution is the globally optimal solution of Hartree-Fock theory. The work extends a previously presented method for closed-shell systems [S. Veeraraghavan and D. A. Mazziotti, Phys. Rev. A 89, 010502(R) (2014))].  For strongly correlated systems the SDP approach provides an alternative to the locally optimized Hartree-Fock energies and densities with a certificate of global optimality. Applications are made to the potential energy curves of C\sub2, CN, Cr\sub2 and NO\sub2.

\end{abstract}

\maketitle

\section{Introduction}

The most widely used approach to obtain the ground-state
Hartree-Fock (HF) energy and density matrix has been to solve the
Euler-Lagrange equations associated with the minimization of the
Hartree-Fock energy. In 1951 Roothaan~\cite{R51, R60} and
Hall~\cite{Hall51} proposed the first self-consistent-field (SCF)
method to solve the Hartree-Fock equations.  However, the method was
soon discovered to converge only for well-behaved cases. Since then
numerous algorithms have been proposed to modify the SCF method to
improve its convergence properties, including
level-shifting~\cite{SH73, Carbo77}, damping~\cite{Hartree57, ZH79}
and direct inversion of the iterative subspace (DIIS)~\cite{P80,
  P82}. DIIS is currently the most popular SCF algorithm because of its
computational efficiency in most cases. Nevertheless, it is not
globally convergent, and in many cases it is known to fail even with a good initial guess.  A method is globally convergent when it converges to a local minimum from any initial guess.  Level-shifting is globally convergent for
a large enough shift parameter~\cite{C00}, but its speed of
convergence decreases as the shift parameter increases.

Since SCF methods do not ensure an energy decrease at each
iteration, a potentially more natural approach to solving the
Hartree-Fock problem is to minimize the Hartree-Fock energy directly
as a function of the density matrix using gradient or Hessian-based
methods.  In 1956 McWeeny~\cite{Mcweeny56} proposed
direct-minimization methods~\cite{Fletcher64, Hillier70, Igawa75,
Seeger76, Camp81, Stanton81, Bacskay81}, but they have not yet found
wide applicability either due to their slow convergence or
prohibitive cost. Recently, a combination of the monotonic energy
decrease property of direct-minimization methods and the speed of
SCF methods was achieved in the relaxed-constraints
algorithms~\cite{CL00, KSC02} and trust-region methods~\cite{T04,
FMM04, FMM06} which are both globally convergent and efficient.
However, none of these methods can certify that the computed
solution is the global minimum.

Convex minimization problems possess the attractive property that the existence of a local minimum implies that it is the global minimum. Semidefinite programs (SDP) are a class of convex optimization problems in which a linear function of a positive semidefinite matrix is optimized subject to linear constraints. For an $N$-electron system the minimization of the ground-state energy as a functional of the two-electron reduced density matrix (2-RDM) subject to $N$-representability constraints~\cite{RDM07, CB, Nrep2} has been expressed as an SDP~\cite{RDM07, E79, ME01, N01, M02}, and the resulting variational 2-RDM method~\cite{RDM07, M12, G75, E79, ME01, N01, M02, Z04, M04, *M06, C06, S10, B12, Nrep2} in conjunction with large-scale SDP solvers~\cite{M04, M11} has been applied to computing directly the 2-RDMs of strongly correlated systems including molecules like polyaromatic hydrocarbons~\cite{Kenley} and firefly luciferin~\cite{firefly} as well as quantum dots~\cite{Adam08}, quantum phase transitions~\cite{Greg06}, and one- and two-dimensional spin models~\cite{Jeff06, B12}.  We recently presented two SDP algorithms that yield upper and lower bounds on the ground-state energy from Hartree-Fock theory. Here we extend these algorithms to treat restricted open-shell Hartree-Fock (ROHF) theory.  While wave function approaches to Hartree-Fock theory yield an upper bound to the Hartree-Fock energy, we derive a semidefinite relaxation of Hartree-Fock theory that yields a rigorous lower bound on the Hartree-Fock energy.  In the lower-bound SDP algorithm the idempotency constraint on the one-electron density matrix is relaxed.  We also develop an upper-bound algorithm in which Hartree-Fock theory is cast as an SDP with a nonconvex constraint on the rank of the matrix variable.  Whenever the upper and lower bounds are equal to each other, they provide a certificate of global optimality to the obtained solution.

To illustrate these algorithms, we apply them to computing the symmetry-broken spin-restricted Hartree-Fock potential energy curves for C\sub2, CN, Cr\sub2 and NO\sub2. This problem is challenging because there are multiple local minima of different spatial symmetries on the potential energy surfaces.  Traditional methods that solve the Euler-Lagrange equations often converge to minima higher in energy than the global minimum.  In principle, \v{C}\'{\i}\v{z}ek-Paldus stability analysis can be applied to locate multiple solutions to the Euler-Lagrange equations, but it can be computationally expensive and it cannot determine whether a local solution is also a global solution.  We find that the upper-bound SDP algorithm consistently converges to the lowest energy solutions and that the lower-bound SDP algorithm generates a tight lower bound.  Neither the upper- or lower-bound SDP algorithm relies on the quality of the initial guess for the density matrix, and in all SDP calculations presented here the initial density matrix is equated to a matrix whose elements are formed by a random number generator.  When the upper and lower bounds are equal, the SDP algorithms provide a certificate of global optimality for the Hartree-Fock solution.  The energetically degenerate symmetry-broken solutions are important because they can be combined convexly into an ensemble density matrix that not only has the desired molecular symmetry but yields a size-consistent energy.  All of the Hartree-Fock solutions that we obtain using the SDP approach are restricted, i.e. $\langle S^2\rangle$ has exactly the correct expectation value.  In many strongly correlated cases, as shown in Section~\ref{sec:apps}, employing the symmetry-broken Hartree-Fock wave function as a reference in single-reference correlation methods like coupled cluster singles-doubles improves the correlated solution.

\section{Theory}

\subsection{Canonical Hartree-Fock Theory}

The quantity that we have been discussing as the density matrix can
be more precisely called the one-particle reduced density matrix
(1-RDM), which we denote as $\one D$. The Hartree-Fock problem for
an $N$-electron system in an orthonormal basis of rank $r$ is
typically expressed as the following minimization problem over the
set of Hermitian matrices ($\mathbb{H}^r$)

\begin{align}
  \operatorname*{minimize}_{\one D \in \mathbb{H}^r}
  & \hspace{0.25cm} E_{\rm HF}(\one D) \\
  \operatorname*{subject\,to}
  & \Tr(\one D) = N \label{d1tr} \\
  & \one D^2 = \one D \label{idem}
\end{align}
The self-consistent-field Hartree-Fock method for an $N$-electron
system iteratively solves a system of Euler-Lagrange equations for a
stationary point. The stationary point yields a ground-state
Hartree-Fock energy and a set of $N$ occupied orbitals. The computed
Hartree-Fock energy is not guaranteed to be the global-energy minimum.
From the perspective of reduced density matrices
(RDMs)~\cite{RDM07,CB}, we can understand the self-consistent-field
method as iteratively checking extreme points of the set of 1-RDMs for
satisfaction of the Euler-Lagrange equations where each extreme point
corresponds to a 1-RDM with a Slater-determinant
preimage~\cite{Nrep,H78,Nrep2}. The set of extreme 1-RDMs (those with
an $N$-electron Slater determinant as a preimage) can be characterized
by the idempotency constraint in Eq.~(\ref{idem}).

\subsection{SDP Hartree-Fock Theory}

\subsubsection{Convex relaxation}

The optimization of the Hartree-Fock energy over the set of extreme
1-RDMs can be replaced without approximation by an optimization over
the larger (and convex) set of $N$-representable 1-RDMs (those with
any $N$-electron wave function as a preimage)~\cite{Lieb81, C00}:
\begin{align}
\operatorname*{minimize}_{\one D, \one Q \in \mathbb{H}^r_+}
& \hspace{0.25cm} E_{\rm HF}(\one D) \label{eq:rdm} \\
\operatorname*{subject\,to}
& \Tr(\one D) = N \label{d1tr2} \\
& \one D + \one Q = I \label{d1q1}
\end{align}
where $E_{\rm HF}$ is the following quadratic function of the 1-RDM:
\begin{align}
E_{\rm HF}(\one D) & = \sum_{ij}^r \one K^i_j \one D^i_j +
\sum_{ijkl}^r \one D^i_k \two
V^{ik}_{jl} \one D^j_l \\
\one K^i_j & = \langle i| {\hat h} |j \rangle \\
\two V^{ik}_{jl} & =  \frac{1}{2}(\langle ij|kl \rangle - \langle
ij|lk \rangle). \label{eq:v2}
\end{align}
The one-electron Hamiltonian operator ${\hat h}$ contains the kinetic
energy operator and electron-nuclei potential, $\langle ij|kl \rangle$
represents the electron-electron repulsion integrals, and the indices
$i$, $j$, $k$, and $l$ denote the orbitals in the one-electron basis
set of rank $r$. The notation $\one D, \one Q \in \mathbb{H}^r_+$,
equivalent to $\one D \succeq 0$ and $\one Q \succeq 0$, indicates
that both the 1-particle RDM $\one D$ and the 1-hole RDM $\one Q$ are
contained in the set of $r \times r$ Hermitian positive semidefinite
matrices.

The reduced-density-matrix formulation of Hartree-Fock theory can be
recast as a convex semidefinite program by embedding the quadratic
product of 1-RDMs in $E_{\rm HF}$ in a higher dimensional
(two-electron) matrix $\two M \in \mathbb{H}^{r^2}_+$. Rewriting
$E_{\rm HF}$ as a linear functional of $\two M$
\begin{align}
 E(\one D, \two M) & = \Tr(\one K \one D) + \Tr(\two V \two M),
\end{align}
we can relax the non-convex Hartree-Fock optimization to a convex
semidefinite program:
\begin{align}
\operatorname*{minimize}_{\one D, \one Q \in \mathbb{H}^{r}_+, \two
M \in \mathbb{H}^{r^2}_+}
 & E(\one D, \two M) \label{eq:minM} \\
\operatorname*{subject\,to} \hspace{0.5cm}
& \Tr(\one D) = N  \\
& \Tr(\two M) \le N  \\
& \one D + \one Q = I  \\
& \sum_{j=1}^r \two M^{ik}_{jj} = N \one D^i_k . \label{eq:conM}
\end{align}
The solution of this SDP relaxation yields a lower bound to the
Hartree-Fock energy.  Because the constraints on the matrix $\two M$
are minimal, this convex SDP formulation will typically yield
energies that are significantly below the Hartree-Fock energy. To
reproduce Hartree-Fock, further constraints on $\two M$ are
required.

\subsubsection{Upper-bound SDP algorithm}

Two separate sets of additional conditions on the matrix $\two M$
that yield upper and lower bounds on the Hartree-Fock energy,
respectively, will be considered.  The first set of constraints,
yielding the upper bound, consists of a single rank constraint
\begin{equation}
\label{eq:rr} \rank{\left(\two M\right)} = 1 .
\end{equation}
The $\two M \in \mathbb{H}^{r^2}_+$ matrix with its rank-one
constraint and the contraction constraint in Eq.~(\ref{eq:conM}), we
can show, is a tensor product of two identical 1-RDMs
\begin{align}
\label{eq:mrr} \two M^{ik}_{jl} & = \one D^i_k \one D^j_l .
\end{align}
It follows that the solution of the optimization program in
Eqs.~(\ref{eq:minM}-\ref{eq:conM}) with the rank constraint in
Eq.~(\ref{eq:rr}) is equivalent to the solution of the RDM
formulation of Hartree-Fock theory in
Eqs.~(\ref{eq:rdm}-\ref{eq:v2}).  We have mapped Hartree-Fock theory
exactly onto a rank-constrained semidefinite program (rc-SDP
HF)~\cite{DB}. The rank-constrained semidefinite program is convex
except for the rank constraint; the nonconvexity of the
Hartree-Fock energy functional in the RDM formulation has been
transferred to the rank restriction in the SDP formulation. Because
of the rank constraint, the solution of rc-SDP HF is not necessarily
a global solution, meaning that the solution can be a local minimum
in the Hartree-Fock energy and hence, an upper bound on the global
energy minimum. Unlike traditional formulations of Hartree-Fock
theory, however, rc-SDP HF optimizes the 1-RDM over the convex set
of $N$-representable 1-RDMs, and in practice, we find that this
difference makes it much more robust than traditional formulations
in locating the global solution.

\subsubsection{Lower-bound SDP algorithm}

\label{sec:lb}

The second set of conditions, yielding a lower bound, consists of
four constraints including
\begin{equation}
\sum_{j=1}^r \two M^{ij}_{jk} = \one D^i_k \label{lb}
\end{equation}
and three additional constraints from permuting the indices $i$ and
$j$ and/or $j$ and $k$ symmetrically.  These convex conditions are a
relaxation of the idempotency of the 1-RDM. They are necessary but
not sufficient for the idempotency of the 1-RDM at the Hartree-Fock
solution, and hence, optimization of the SDP program in
Eqs.~(\ref{eq:minM}-\ref{eq:conM}) with these additional constraints
(lb-SDP) is an SDP relaxation of the reduced-density-matrix
formulation of Hartree-Fock theory in
Eqs.~(\ref{eq:rdm}-\ref{eq:v2}). The lb-SDP method yields a {\em
lower bound on the energy} from the global Hartree-Fock solution. In
practice, this lower bound is found to be quite tight, and in some
cases it agrees exactly with the global Hartree-Fock solution. If
lb-SDP produces a 1-RDM solution that is idempotent, then that
solution is the global Hartree-Fock solution. Furthermore, when the
upper and lower bounds from rc-SDP and lb-SDP agree, we have a
guaranteed certificate that these computed bounds correspond to the
global energy minimum of Hartree-Fock theory.

\subsubsection{Closed- and open-shell spin restriction}

We have formulated rc-SDP HF and lb-SDP in the spin-orbital basis
set. To perform RHF and ROHF calculations in a spatial-orbital basis
set, one needs to take into account the spin structure of the
Hamiltonian and density matrices. For any RHF or ROHF calculation on an
$N$-electron system, $\one D$, $\two M$, $\one K$ and $\two V$ will have
the following block structures:
\begin{align}
  \one D & =
  \begin{bmatrix}
    \one D_\alpha & 0 \\
     0 &  \one D_\beta \\
  \end{bmatrix} &
  \two M & =
  \begin{bmatrix}
    \two M_{\alpha \alpha} & \two M_{\alpha \beta} \\
     \two M^t_{\alpha \beta} &  \two M_{\beta \beta} \\
  \end{bmatrix} & \\
\notag \\
\one K & =
  \begin{bmatrix}
    \one K_\alpha & 0 \\
     0 &  \one K_\alpha\\
  \end{bmatrix} &
\two V & =
  \begin{bmatrix}
    \two V_{\alpha \alpha} & \two V_{\alpha \beta} \\
     \two V_{\alpha \beta} &  \two V_{\alpha \alpha} \\
  \end{bmatrix} . &
\end{align}
For RHF the two $\one D$ blocks and the four blocks of $\two M$ are
identical. Therefore the only necessary modifications in rc-SDP HF and
lb-SDP are replacing $N$ by $N/2$, $r$ by $r/2$ and rewriting $E$ as
follows:
\begin{multline}
  E(\one D_\alpha, \two M_{\alpha \beta}) \\ = 2\Tr(\one K_\alpha\one D_\alpha) + \Tr(\two
  \mathcal{V} \two M_{\alpha \beta})
  \label{EM-RHF}
\end{multline}
\begin{align}
  \two \mathcal{V} & = 2(\two V_{\alpha \alpha}
  + \two V_{\alpha \beta}) \\
  \two \mathcal{V}^{ik}_{jl} & = 2\langle ij|kl \rangle - \langle ij|lk \rangle .
\end{align}
For ROHF the $\alpha$ and $\beta$ blocks of $\one D$ are not identical but
(assuming $N_\alpha > N_\beta$) because the spatial orbitals of paired electrons are required to be the same, the row space of $\one D_\beta$ is a
subset of the row space of $\one D_\alpha$.  In order to
enforce this relation, $\one D$ is divided into closed-shell and
open-shell blocks which are $\one D_c = \one D_\beta$ and $\one D_{o}
= \one D_\alpha - \one D_\beta$ respectively. With these blocks $E$
and SDP ROHF can be rewritten as follows:
\begin{multline}
  E(\one D_c, \one D_o,  \two M) \\ = 2\Tr(\one K_\alpha \one D_c) + \Tr(\one K_\alpha \one D_o) +
  \Tr(\two V \two M) \label{EM-ROHF}
\end{multline}
\begin{align*}
  \operatorname*{minimize}_{\one D_c, \one D_o, \one Q \in \mathbb{H}^{r}_+,
    \two M \in \mathbb{H}^{r^2}_+}
  & E(\one D_c, \one D_o,  \two M) \\
  \operatorname*{subject\,to} \hspace{0.9cm}
  & \Tr(\one D_c) = N_\beta  \\
  & \Tr(\one D_o) = N_\alpha - N_\beta  \\
  & \one D_c + \one D_o + \one Q = I  \\
  & \sum_{j=1}^r \two M^{i\alpha k\alpha}_{j\alpha j\alpha} = N_\alpha \left( \one {D_c}^i_k + \one {D_o}^i_k \right) &   \\
  & \sum_{j=1}^r \two M^{i\beta k\beta}_{j\beta j\beta} = N_\beta \one {D_c}^i_k &   \\
  & \sum_{j=1}^r \two M^{i\alpha k\alpha}_{j\beta j\beta} = N_\beta \left( \one {D_c}^i_k + \one {D_o}^i_k \right) &   \\
  & \sum_{j=1}^r \two M^{i\beta k\beta}_{j\alpha j\alpha} = N_\alpha \one {D_c}^i_k & \\ .
\end{align*}
For rc-SDP ROHF, the only additional constraint is on the rank, as in the spin-orbital formulation
\begin{align}
  \rank{\left( \two M \right)} = 1 .
\end{align}
For lb-SDP, the set of four constraints described earlier in
Eq.~(\ref{lb}) have to be enforced for all the four blocks of $M$ as
follows:
\begin{align}
\sum_{j=1}^r \two M^{i\alpha j\alpha}_{j\alpha k\alpha} & = \left( \one {D_c}^i_k + \one {D_o}^i_k \right) \\
\sum_{j=1}^r \two M^{i\beta j\beta}_{j\beta k\beta} & = \one {D_c}^i_k \\
\sum_{j=1}^r \two M^{i\alpha j\alpha}_{j\beta k\beta} & = \one {D_c}^i_k \\
\sum_{j=1}^r \two M^{i\beta j\beta}_{j\alpha k\alpha} & = \one {D_c}^i_k .
\end{align}
Although formally there are 16 constraints in all, 6 of them are
redundant due to the Hermiticity of $M$ resulting in 10 linearly
independent constraints.

\section{Applications}

\label{sec:apps}

To illustrate the rc-SDP and lb-SDP methods, we apply them to
computing the dissociation curves for C\sub2, CN, Cr\sub2 and
NO\sub2.

\subsection{Methodology}

The GAMESS electronic structure package is used to perform self-consistent-field Hartree-Fock calculations (SCF HF with DIIS) and coupled cluster singles doubles (CCSD)~\cite{PB82,BM07, CCSD} calculations. The rc-SDP and lb-SDP are solved using the SDP solver RRSDP~\cite{M04}. Since DIIS is the standard accelerator for SCF HF calculations, we compare rc-SDP HF results with DIIS results. Both rc-SDP HF and DIIS methods are performed without enforcing a specific spatial symmetry.  The DIIS solution at the internuclear distance $R'$ where $R'$ is differentially larger than the distance $R$ is obtained by using the DIIS solution at $R$ as an initial guess.

The SDP solver RRSDP imposes the semidefinite constraint on each
matrix $M$ through the factorization $M = RR^T$.  For rc-SDP HF, the
rank-one constraint on $\two M$ is readily enforced by defining $R$ to
be a rectangular $r \times 1$ matrix. Scaling of RRSDP~\cite{M04} is
determined by the $RR^T$ matrix multiplication for the largest matrix
block, which is $\two M$ for both rc-SDP and lb-SDP. For rc-SDP the
rank of $\two M$ is one, and hence, the matrix multiplication scales
approximately as $r^4$. For lb-SDP the rank of $\two M$ scales as $r$
after applying the bound on the maximum rank from Pataki~\cite{P98}
and Barvinok~\cite{B95}, and hence, the matrix multiplication scales
approximately as $r^5$.

\begin{figure}[htp!]
\includegraphics[scale=0.7]{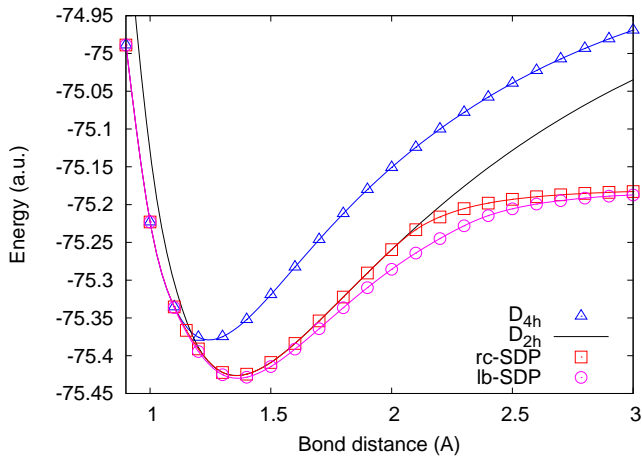} \\
\caption{The ground-state restricted Hartree-Fock energies from rc-SDP
  and DIIS and the lower bound from lb-SDP are shown as functions of
  the C-C internuclear distance $R$. When the energies from
  rc-SDP and lb-SDP agree, the solution from rc-SDP is guaranteed to
  be the global solution of Hartree-Fock theory. While DIIS locates a
  D\sub{4h} and a D\sub{2h} solution depending on the initial guess
  used, rc-SDP locates the energetically lowest solution which has a
  different symmetry for different internuclear distances. Although
  the rc-SDP solution has C\sub{i} symmetry for $R > 2$~\AA\ it has a
  qualitatively correct shape for dissociation. By 3~\AA\ the
  D\sub{2h} and C\sub{i} solutions differ by
  0.148~a.u. (92.8~kcal/mol). }
\label{C2-HF}
\end{figure}

\subsection{C\sub{2} stretch}

Because the C\sub2 molecule has many low-lying excited states, it is a
significantly multireferenced system even at equilibrium, which makes
it a challenging system for both Hartree-Fock and correlation energy
calculations~\cite{GM05}. Figure \ref{C2-HF} shows various RHF energy
curves and the lower bound in the 6-31G\textsuperscript* basis set as
a function of the C-C bond distance. The D\sub{4h} and D\sub{2h}
curves~\cite{AS04}, generated by seeding the scan of the potential
energy surface at two different values of $R$ were obtained using
DIIS. The rc-SDP curve corresponds to D\sub{4h} for $R < 1.1$~\AA,
C\sub{s} for $1.1$~\AA\ $< R < 1.2$~\AA, D\sub{2h} for $1.2$~\AA\ $< R < 2$~\AA, and to C\sub{i} for $R > 2$~\AA. The rc-SDP curve bifurcates from the D\sub{4h} curve and joins the D\sub{2h} curve without ever being
non-differentiable. For $R < 1.5$~\AA\ and $R > 2.9$~\AA\, the lb-SDP solution is lower than the rc-SDP solution by less than 0.005 a.u. thereby
certifying it to be the global minimum within that threshold. For the
intermediate region, rc-SDP likely continues to give the globally
optimal curve although we do not have a formal mathematical guarantee.

\subsection{Cr\sub{2} stretch}

\begin{figure}[htp!]
\includegraphics[scale=0.7]{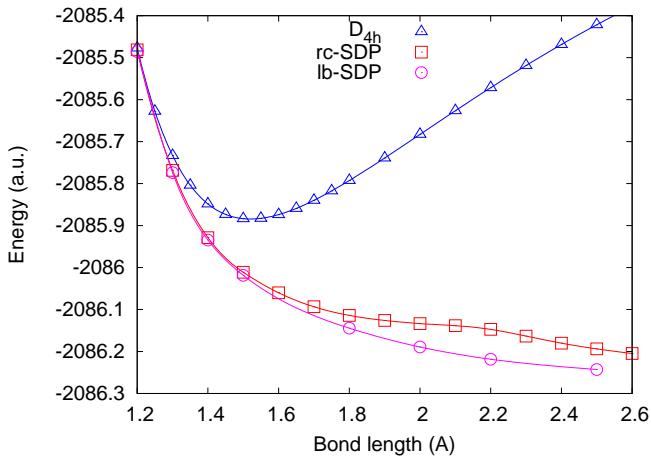} \\
  \caption{The Hartree-Fock energies from rc-SDP and DIIS and the lower bound from lb-SDP are shown as functions of the Cr-Cr internuclear distance $R$. DIIS obtains a D\sub{4h} solution on seeding from smaller values of $R$ to larger ones.  While for $R > 1.2$~\AA\ (shown) the rc-SDP solution has C\sub{2} symmetry, for $R < 1.2$~\AA\ (not shown) it smoothly joins the D\sub{4h} curve of DIIS.  Even in the equilibrium region, for $R$~=~1.5~\AA\ the D\sub{4h} energy is 0.127 a.u. (80 kcal/mol) higher than the C\sub{2} energy which is certified by lb-SDP to be globally optimal within 0.008 a.u.  By 2.5~\AA\ the energy difference increases to 0.772~a.u. (485~kcal/mol) where the C\sub2 solution is globally optimal within 0.05~a.u.}
\label{Cr2-HF}
\end{figure}

\begin{figure}[htp!]
\includegraphics[scale=0.7]{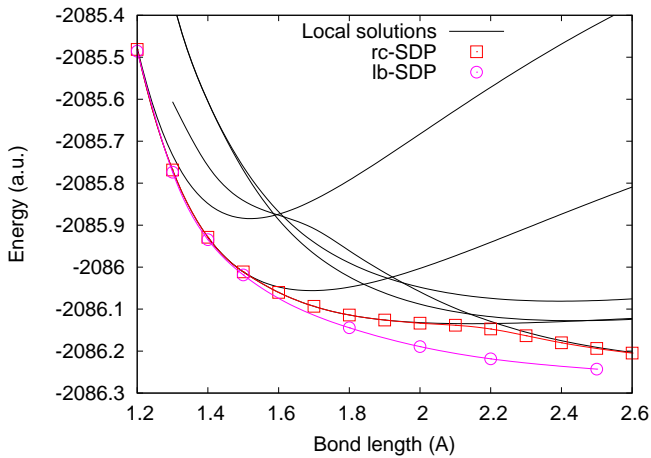} \\
\caption{All the local Hartree-Fock solutions found are shown in order
  to provide a visual depiction of the substantial multireference
  correlation that exists in Cr\sub2 and the concomitant difficulty in
  obtaining global Hartree-Fock solutions.}
\label{Cr2-HF-local}
\end{figure}

Cr\sub2 is known to be an extremely challenging molecule to describe
correctly by \textit{ab initio} electronic structure
theory~\cite{Good85, Roos03, Celani04, Muller09, And94, Bausch94,
  Roos95, Zgid09, Angeli06}. Figure \ref{Cr2-HF} shows the HF energy
in the valence triple-zeta (TZV)~\cite{TZV} basis set as a function of
the Cr-Cr distance. The large number of HF solutions that are energetically close to each other, shown in Fig.~\ref{Cr2-HF-local}, provides
a novel characterization of the substantial multireference correlation
in Cr\sub2. The number of energetically close HF solutions is comparable in the STO-6G basis, indicating that this feature is not significantly dependent upon the basis set.  The solution found by DIIS has D\sub{4h} symmetry for all $R$ whereas the solution found by rc-SDP has D\sub{4h} symmetry for $R < 1.2$~\AA\ and C\sub2 symmetry for $R > 1.2$~\AA. Although the rc-SDP
solution is symmetry-broken for $R > 1.2$~\AA, it is globally optimal within the bound provided by lb-SDP. Further verification of the rc-SDP solutions being HF minima is provided by the fact that DIIS is able to obtain them
when they are employed as initial guesses.

\begin{figure}[htp!]
\includegraphics[scale=0.7]{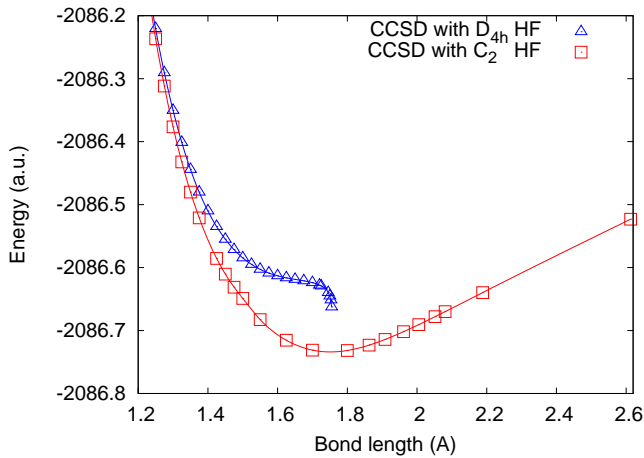} \\
  \caption{The potential energy curves of Cr$_{2}$ from CCSD with the D\sub{4h} and C\sub{2} Hartree-Fock wave functions are compared. The CCSD method applied with the D\sub{4h} reference yields an unphysical curve. In contrast, CCSD with the C\sub2 reference yields a physically realistic dissociation curve. Consequently, the energy divergence from CCSD can be attributed to the D\sub{4h} reference wave function.}
\label{Cr2-CCSD}
\end{figure}

Changes in Hartree-Fock energies and densities can impact
correlation energy calculations in two ways: (1) any change in the
Hartree-Fock energy changes the correlation energy by its very
definition and (2) any change in the Hartree-Fock density (or the
Hilbert space spanned by the molecular orbitals) changes the
reference wave function employed in many-electron correlation
methods including coupled cluster~\cite{PB82,BM07,CCSD} and
parametric RDM methods~\cite{M08,M10}. In this and other examples considered, rc-SDP helps to identify symmetry-broken Hartree-Fock solutions that often generate improved CCSD solutions.  While rc-SDP may identify a piecewise smooth potential energy surface, each piece can be analytically continued to generate a smooth Hartree-Fock surface from which a smooth CCSD surface can be computed.  Cr\sub2 is known to be extremely challenging for single-reference methods like coupled
cluster theories~\cite{Cr2-CCSD}. Figure~\ref{Cr2-CCSD} explores the
effect of using the global symmetry-broken C\sub2 solution rather
than the local D\sub{4h} solution as the reference wave function in
CCSD. The results in Fig.~\ref{Cr2-CCSD} show that much of the
failure noted in the literature can be attributed to
the D\sub{4h} reference wave function rather than CCSD.  While CCSD
with the D\sub{4h} reference diverges beyond 1.7 \AA\ , CCSD with
the C\sub2 reference at least yields a physically realistic
dissociation curve for the ground state.


\begin{figure}[htp!]
\includegraphics[scale=0.7]{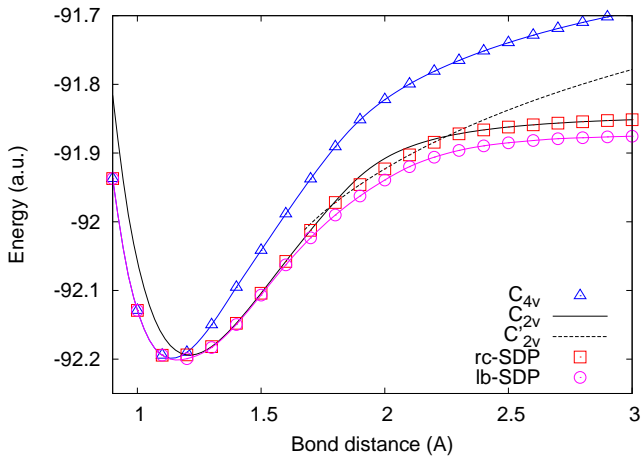} \\
  \caption{The Hartree-Fock energies from rc-SDP and DIIS and the lower
  bound from lb-SDP are shown as functions of the C-N internuclear
  distance. In spite of there being multiple solutions, rc-SDP
  successfully obtains the lowest energy solution for all internuclear
  distances. By 3~\AA\ the difference between the C\sub{4v} and
  C\sub{2v} solutions is 0.156~a.u. (98~kcal/mol).}
\label{CN-HF}
\end{figure}

\subsection{CN stretch}

The CN radical is of astrophysical interest due its presence in the
interstellar medium~\cite{C80}. Although its low-lying excited
states~\cite{KWHC88} enhance its utility as a optical probe
~\cite{MO69, WB74} for studying various properties, they also make it
a multireferenced system. Figure \ref{CN-HF} shows various HF energy
curves and the lower bound in the cc-pVDZ basis~\cite{D89} plotted as
a function of the C-N internuclear distance $R$. The C\sub{4v},
C\sub{2v} and C'\sub{2v} curves were obtained using DIIS. The rc-SDP
curve corresponds to C\sub{4v} for $R < 1.2$ \AA\ and to C\sub{2v} for
all other values of $R$.
The rc-SDP method manages to obtain the lowest curve among three
different Hartree-Fock curves for all values of $R$. As is evident
from the figure, the ground-state Hartree-Fock curve is a piecewise
defined function of the three Hartree-Fock curves with there being
three points of non-smoothness where the curves intersect at $R$ =
1.2, 1.8 and 2.25 \AA\ . If DIIS is given the H\"uckel guess, it
converges to different curves in different regions which are not
always the lowest solutions for those regions. If DIIS is used to
generate the curve from left to right with the solution for a smaller
bond distance being the guess for a larger bond distance, only the
C\sub{4v} curve is obtained. Consistent generation of the C\sub{4v}
curve might be the reason why the C\sub{2v} and C'\sub{2v} curves have
not been previously reported~\cite{TO04}. The fact that the lb-SDP
curve is never lower than rc-SDP by more than 0.006~a.u. for $R \le
1.6$~\AA\ provides a certificate of global optimality for those rc-SDP
points within that threshold. After 1.6~\AA\ rc-SDP likely continues
to give the globally optimal curve although we do not have a formal
mathematical guarantee. This is corroborated by the fact that the
rc-SDP (and C\sub{2v}) curve is size consistent, meaning that it is
asymptotically equal to exactly the sum of Hartree-Fock energies of
doublet N and singlet C. The energy of doublet N and singlet C is the
same from both rc-SDP and DIIS and certified by lb-SDP to be globally optimal within 0.002 a.u. and 0.0004 a.u. respectively.

It is also worth noting that rc-SDP (and the C\sub{2v} curve) does not
dissociate CN into quadruplet N and triplet C in spite of them being
lower in energy than doublet N and singlet C respectively. This is not
due to convergence to a local minimum but instead is due to the
inability of `restricted' orbitals in ROHF (and RHF) to dissociate a
molecule into fragments which have spatially separated electron
pairs. This is a consequence of using the same spatial orbital to
describe electron pairs. Since (doublet) CN has one unpaired electron,
ROHF can describe its dissociation into fragments which have a total
of one unpaired electron at most which is why it dissociates CN into
doublet N and singlet C.

\begin{figure}[htp!]
\includegraphics[scale=0.7]{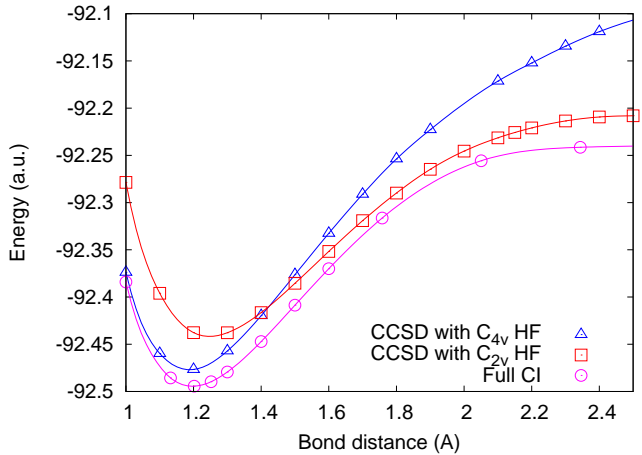} \\
  \caption{The potential energy curves of CN from CCSD with the C\sub{4v} and C\sub{2v} Hartree-Fock references are shown as functions of the C-N internuclear distance. Like the C\sub{4v} and C\sub{2v} Hartree-Fock solutions, the C\sub{4v} and C\sub{2v} CCSD solutions provide good descriptions at equilibrium and dissociation, respectively.}
\label{CN-CCSD}
\end{figure}

Figure \ref{CN-CCSD} shows the CCSD curves obtained using the
C\sub{4v} and C\sub{2v} Hartree-Fock solutions of which the latter was identified using rc-SDP. As is evident from the
figure, although the CCSD with the C\sub{4v} reference curve is lower
in energy for R~$\leq$~1.4~\AA\ , it rises rapidly after that point as
it dissociates into charged species. The CCSD with the C\sub{2v}
reference curve is lower in energy after 1.4 \AA\ and gives a
qualitatively correct representation of dissociation into neutral
species. It is also worth noting that the C\sub{4v} and C\sub{2v}
Hartree-Fock curves cross at 1.2 \AA, which is before the
corresponding CCSD curves cross.

\begin{figure}[htp!]
\includegraphics[scale=0.7]{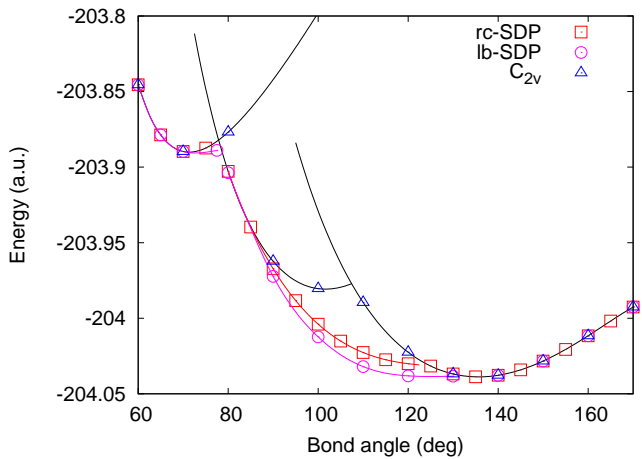} \\
\caption{The Hartree-Fock energies from rc-SDP and DIIS (all of which
  correspond to C\sub{2v} symmetry) and the lower bound from lb-SDP
  are shown as functions of the ONO angle. Despite there being
  multiple C\sub{2v} solutions, rc-SDP always finds the energetically
  lowest solution. Furthermore, lb-SDP certifies global optimality for
  the solutions $< 70^\circ$ and $> 135^\circ$.}
\label{NO2-HF}
\end{figure}

\subsection{NO\sub{2} bend}

The NO\sub2 radical is known to have a complicated, extensively
studied photochemistry~\cite{WW10}. It has a conical intersection
between the ground and first excited states~\cite{Worner11,
  Kraus12}. Figure \ref{NO2-HF} shows various HF energy curves
(C\sub{2v} symmetry) and the lower bound in the cc-pVDZ
basis~\cite{D89} plotted as a function of the O-N-O angle ($\theta$)
for a symmetric configuration with a N-O bond length of \mbox{ 1.197
  \AA\ .} Despite the existence of multiple HF minima which are
energetically close, rc-SDP obtains the lowest minimum for all bond
angles. For $\theta \leq 70^\circ , \theta \geq 135^\circ$ lb-SDP
certifies the C\sub{2v} curve from both DIIS and rc-SDP to be globally
optimal. Furthermore, since lb-SDP is never lower than rc-SDP by more
than 0.01~a.u., the entire rc-SDP curve is globally optimal within
that threshold. Although the rc-SDP curve for $85^\circ \leq \theta
\leq 120^\circ$ corresponds to a saddle point on the complete NO\sub2
potential energy surface (as a function of the two bond lengths in
addition to the bond angle) it is indeed the global minimum (within
0.01~a.u.) for the fixed values of N-O bond lengths used in the
calculation.

\section{Discussion}

An RDM formulation of Hartree-Fock theory, based on semidefinite programming, has been presented that yields upper and lower bounds on the Hartree-Fock solution.  When these bounds are equal, they provide a certificate guaranteeing the globally optimal Hartree-Fock solution.  As electrons become more strongly correlated, methods for Hartree-Fock based on the self-consistent-field approach like DIIS converge to stationary points of Hartree-Fock theory with potentially non-global energies and densities. In most instances we have been able to certify global optimality for known solutions of Hartree-Fock theory for the first time. Although there are methods to determine whether the obtained stationary point is a local minimum, maximum, or saddle point~\cite{CP67,SP77}, our approach is unique in that it certifies global optimality.

Semidefinite relaxation of Hartree-Fock theory, which we derived in Section~\ref{sec:lb}, yields a rigorous lower bound on the Hartree-Fock energy. In contrast, wave function approaches to Hartree-Fock theory, such as the traditional optimization of a Slater determinant, yield upper bounds on the Hartree-Fock energy.  Minimization of the electronic energy with respect to the orbitals of a Slater determinant generates a local stationary point which may or may not be the global minimum. Higher derivatives, such as those found in stability analysis, can be employed to search for additional local minima, each of which provides an upper bound on the energy of the global minimum.  While not previously developed, the lower-bound approach enables us in many cases to certify that a solution is the global minimum of Hartree-Fock theory. If the 1-RDM obtained by the lower-bound SDP algorithm is idempotent, then the 1-RDM and its associated energy represent the global solution to the Hartree-Fock calculation. Furthermore, even if the 1-RDM is not idempotent, agreement of the lower-bound energy with the upper-bound energy from either a traditional wave function-based Hartree-Fock calculation or an SDP-based upper-bound calculation guarantees that the computed energy is the global minimum. As shown in the theory section, an upper bound to the Hartree-Fock energy can be computed through a rank-constrained SDP in which a nonconvex rank constraint is added to the optimization. Importantly, this upper-bound formulation shows that Hartree-Fock theory is convex except for the presence of the rank constraint.

Symmetry breaking and restoration can be employed to capture correlation effects at a lower computational cost~\cite{Scu13}. Recently, they have been employed in variational quasi-particle theory~\cite{SJH11} to compute the ground-state energies from an antisymmetrized geminal power wave function\cite{M00, *M01b} at an $r^4$ computational cost where $r$ is the number of orbitals.  In the presence of strong electron correlation the lowest energy solutions of Hartree-Fock theory can be spatially symmetry broken. We employ the SDP methods to distinguish the global solution from multiple local solutions of different spatial symmetries.  The symmetry breaking generates multiple wave functions at the global minimum that are energetically degenerate.  In the present case symmetry restoration can be accomplished by two methods. First, the full molecular symmetry can be reestablished by taking the ensemble of the energetically degenerate symmetry-broken solutions. The ensemble nature of the ground-state density matrix is a consequence of pursuing a mean-field description of the strongly correlated system. Second, a linear combination of the symmetry-broken solutions can be taken to generate a wave function by a non-orthogonal configuration interaction~\cite{NOCI, NOCI2}.  In the second approach the degenerate Hartree-Fock solutions become entangled to form a pure density matrix composed of a single correlated wave function. While in the present work we pursue the first approach, the second approach provides insight into how the different symmetry-broken solutions of Hartree-Fock theory contribute information to the correlated ground-state wave function.

Direct computation of the two-electron reduced density matrix (2-RDM) has been previously accomplished by minimizing the energy as a functional of the 2-RDM subject to $N$-representability conditions~\cite{RDM07, M12, G75, E79, ME01, N01, M02, Z04, M06, C06, S10, B12, Nrep2}.  Constrained optimization is performed by SDP.  Although we have taken a different path in the derivation of the SDP algorithms for Hartree-Fock theory, they can be viewed within variational 2-RDM theory as the addition of further constraints on the 2-RDM to ensure that it represents the mean-field (or Hartree-Fock) limit.  The upper-bound algorithm requires a nonconvex rank constraint, and the lower-bound algorithm requires relaxed idempotency conditions.  As described above, the combination of the upper-bound and lower-bound SDP algorithms provides a mechanism for certifying the global minimum of Hartree-Fock theory.  We have shown that global solutions are useful for seeding either wave function or reduced density matrix methods for describing strongly correlated quantum systems.  The SDP-based restricted closed- and open-shell Hartree-Fock method, described here, is directly extendable to an unrestricted Hartree-Fock method, which will be presented elsewhere.

\begin{acknowledgments}

DAM gratefully acknowledges the National Science Foundation, the Army Research Office, and Microsoft Corporation for their generous support.

\end{acknowledgments}


\begin{thebibliography}{0}%
\makeatletter
\providecommand \@ifxundefined [1]{%
 \@ifx{#1\undefined}
}%
\providecommand \@ifnum [1]{%
 \ifnum #1\expandafter \@firstoftwo
 \else \expandafter \@secondoftwo
 \fi
}%
\providecommand \@ifx [1]{%
 \ifx #1\expandafter \@firstoftwo
 \else \expandafter \@secondoftwo
 \fi
}%
\providecommand \natexlab [1]{#1}%
\providecommand \enquote  [1]{``#1''}%
\providecommand \bibnamefont  [1]{#1}%
\providecommand \bibfnamefont [1]{#1}%
\providecommand \citenamefont [1]{#1}%
\providecommand \href@noop [0]{\@secondoftwo}%
\providecommand \href [0]{\begingroup \@sanitize@url \@href}%
\providecommand \@href[1]{\@@startlink{#1}\@@href}%
\providecommand \@@href[1]{\endgroup#1\@@endlink}%
\providecommand \@sanitize@url [0]{\catcode `\\12\catcode `\$12\catcode
  `\&12\catcode `\#12\catcode `\^12\catcode `\_12\catcode `\%12\relax}%
\providecommand \@@startlink[1]{}%
\providecommand \@@endlink[0]{}%
\providecommand \url  [0]{\begingroup\@sanitize@url \@url }%
\providecommand \@url [1]{\endgroup\@href {#1}{\urlprefix }}%
\providecommand \urlprefix  [0]{URL }%
\providecommand \Eprint [0]{\href }%
\providecommand \doibase [0]{http://dx.doi.org/}%
\providecommand \selectlanguage [0]{\@gobble}%
\providecommand \bibinfo  [0]{\@secondoftwo}%
\providecommand \bibfield  [0]{\@secondoftwo}%
\providecommand \translation [1]{[#1]}%
\providecommand \BibitemOpen [0]{}%
\providecommand \bibitemStop [0]{}%
\providecommand \bibitemNoStop [0]{.\EOS\space}%
\providecommand \EOS [0]{\spacefactor3000\relax}%
\providecommand \BibitemShut  [1]{\csname bibitem#1\endcsname}%
\let\auto@bib@innerbib\@empty
\end{thebibliography}%


\begin{thebibliography}{85}%
\makeatletter
\providecommand \@ifxundefined [1]{%
 \@ifx{#1\undefined}
}%
\providecommand \@ifnum [1]{%
 \ifnum #1\expandafter \@firstoftwo
 \else \expandafter \@secondoftwo
 \fi
}%
\providecommand \@ifx [1]{%
 \ifx #1\expandafter \@firstoftwo
 \else \expandafter \@secondoftwo
 \fi
}%
\providecommand \natexlab [1]{#1}%
\providecommand \enquote  [1]{``#1''}%
\providecommand \bibnamefont  [1]{#1}%
\providecommand \bibfnamefont [1]{#1}%
\providecommand \citenamefont [1]{#1}%
\providecommand \href@noop [0]{\@secondoftwo}%
\providecommand \href [0]{\begingroup \@sanitize@url \@href}%
\providecommand \@href[1]{\@@startlink{#1}\@@href}%
\providecommand \@@href[1]{\endgroup#1\@@endlink}%
\providecommand \@sanitize@url [0]{\catcode `\\12\catcode `\$12\catcode
  `\&12\catcode `\#12\catcode `\^12\catcode `\_12\catcode `\%12\relax}%
\providecommand \@@startlink[1]{}%
\providecommand \@@endlink[0]{}%
\providecommand \url  [0]{\begingroup\@sanitize@url \@url }%
\providecommand \@url [1]{\endgroup\@href {#1}{\urlprefix }}%
\providecommand \urlprefix  [0]{URL }%
\providecommand \Eprint [0]{\href }%
\providecommand \doibase [0]{http://dx.doi.org/}%
\providecommand \selectlanguage [0]{\@gobble}%
\providecommand \bibinfo  [0]{\@secondoftwo}%
\providecommand \bibfield  [0]{\@secondoftwo}%
\providecommand \translation [1]{[#1]}%
\providecommand \BibitemOpen [0]{}%
\providecommand \bibitemStop [0]{}%
\providecommand \bibitemNoStop [0]{.\EOS\space}%
\providecommand \EOS [0]{\spacefactor3000\relax}%
\providecommand \BibitemShut  [1]{\csname bibitem#1\endcsname}%
\let\auto@bib@innerbib\@empty
\bibitem [{\citenamefont {Roothaan}(1951)}]{R51}%
  \BibitemOpen
  \bibfield  {author} {\bibinfo {author} {\bibfnamefont {C.~C.~J.}\
  \bibnamefont {Roothaan}},\ }\href {\doibase 10.1103/RevModPhys.23.69}
  {\bibfield  {journal} {\bibinfo  {journal} {Rev. Mod. Phys.}\ }\textbf
  {\bibinfo {volume} {23}},\ \bibinfo {pages} {69} (\bibinfo {year}
  {1951})}\BibitemShut {NoStop}%
\bibitem [{\citenamefont {Roothaan}(1960)}]{R60}%
  \BibitemOpen
  \bibfield  {author} {\bibinfo {author} {\bibfnamefont {C.~C.~J.}\
  \bibnamefont {Roothaan}},\ }\href {\doibase 10.1103/RevModPhys.32.179}
  {\bibfield  {journal} {\bibinfo  {journal} {Rev. Mod. Phys.}\ }\textbf
  {\bibinfo {volume} {32}},\ \bibinfo {pages} {179} (\bibinfo {year}
  {1960})}\BibitemShut {NoStop}%
\bibitem [{\citenamefont {Hall}(1951)}]{Hall51}%
  \BibitemOpen
  \bibfield  {author} {\bibinfo {author} {\bibfnamefont {G.}~\bibnamefont
  {Hall}},\ }\href
  {http://rspa.royalsocietypublishing.org/content/205/1083/541.short}
  {\bibfield  {journal} {\bibinfo  {journal} {Proc. R. Soc. A}\ }\textbf
  {\bibinfo {volume} {205}},\ \bibinfo {pages} {541} (\bibinfo {year}
  {1951})}\BibitemShut {NoStop}%
\bibitem [{\citenamefont {Saunders}\ and\ \citenamefont
  {Hillier}(1973)}]{SH73}%
  \BibitemOpen
  \bibfield  {author} {\bibinfo {author} {\bibfnamefont {V.~R.}\ \bibnamefont
  {Saunders}}\ and\ \bibinfo {author} {\bibfnamefont {I.~H.}\ \bibnamefont
  {Hillier}},\ }\href {\doibase 10.1002/qua.560070407} {\bibfield  {journal}
  {\bibinfo  {journal} {Int. J. Quantum Chem.}\ }\textbf {\bibinfo {volume}
  {7}},\ \bibinfo {pages} {699} (\bibinfo {year} {1973})}\BibitemShut {NoStop}%
\bibitem [{\citenamefont {Carbo}, \citenamefont {Hernandez},\ and\
  \citenamefont {Sanz}(1977)}]{Carbo77}%
  \BibitemOpen
  \bibfield  {author} {\bibinfo {author} {\bibfnamefont {R.}~\bibnamefont
  {Carbo}}, \bibinfo {author} {\bibfnamefont {J.}~\bibnamefont {Hernandez}}, \
  and\ \bibinfo {author} {\bibfnamefont {F.}~\bibnamefont {Sanz}},\ }\href
  {\doibase 10.1016/0009-2614(77)85046-X} {\bibfield  {journal} {\bibinfo
  {journal} {Chem. Phys. Lett.}\ }\textbf {\bibinfo {volume} {47}},\ \bibinfo
  {pages} {581 } (\bibinfo {year} {1977})}\BibitemShut {NoStop}%
\bibitem [{\citenamefont {Hartree}(1957)}]{Hartree57}%
  \BibitemOpen
  \bibfield  {author} {\bibinfo {author} {\bibfnamefont {D.~R.}\ \bibnamefont
  {Hartree}},\ }\href {http://books.google.com/books?id=FEtAAAAAIAAJ} {\emph
  {\bibinfo {title} {The calculation of atomic structures}}},\ Structure of
  matter series\ (\bibinfo  {publisher} {J. Wiley},\ \bibinfo {year}
  {1957})\BibitemShut {NoStop}%
\bibitem [{\citenamefont {Zerner}\ and\ \citenamefont
  {Hehenberger}(1979)}]{ZH79}%
  \BibitemOpen
  \bibfield  {author} {\bibinfo {author} {\bibfnamefont {M.~C.}\ \bibnamefont
  {Zerner}}\ and\ \bibinfo {author} {\bibfnamefont {M.}~\bibnamefont
  {Hehenberger}},\ }\href {\doibase 10.1016/0009-2614(79)80761-7} {\bibfield
  {journal} {\bibinfo  {journal} {Chem. Phys. Lett.}\ }\textbf {\bibinfo
  {volume} {62}},\ \bibinfo {pages} {550 } (\bibinfo {year}
  {1979})}\BibitemShut {NoStop}%
\bibitem [{\citenamefont {Pulay}(1980)}]{P80}%
  \BibitemOpen
  \bibfield  {author} {\bibinfo {author} {\bibfnamefont {P.}~\bibnamefont
  {Pulay}},\ }\href {\doibase 10.1016/0009-2614(80)80396-4} {\bibfield
  {journal} {\bibinfo  {journal} {Chem. Phys. Lett.}\ }\textbf {\bibinfo
  {volume} {73}},\ \bibinfo {pages} {393 } (\bibinfo {year}
  {1980})}\BibitemShut {NoStop}%
\bibitem [{\citenamefont {Pulay}(1982)}]{P82}%
  \BibitemOpen
  \bibfield  {author} {\bibinfo {author} {\bibfnamefont {P.}~\bibnamefont
  {Pulay}},\ }\href {\doibase 10.1002/jcc.540030413} {\bibfield  {journal}
  {\bibinfo  {journal} {J. Comp. Chem.}\ }\textbf {\bibinfo {volume} {3}},\
  \bibinfo {pages} {556} (\bibinfo {year} {1982})}\BibitemShut {NoStop}%
\bibitem [{\citenamefont {Canc\`{e}s}(2000)}]{C00}%
  \BibitemOpen
  \bibfield  {author} {\bibinfo {author} {\bibfnamefont {E.}~\bibnamefont
  {Canc\`{e}s}},\ }\href@noop {} {\emph {\bibinfo {title} {Mathematical Models
  and Methods for Ab Initio Quantum Chemistry}}},\ \bibinfo {edition} {1st}\
  ed.,\ edited by\ \bibinfo {editor} {\bibfnamefont {M.}~\bibnamefont
  {Defranceschi}}\ and\ \bibinfo {editor} {\bibfnamefont {C.~L.}\ \bibnamefont
  {Bris}}\ (\bibinfo  {publisher} {Springer},\ \bibinfo {address} {New York},\
  \bibinfo {year} {2000})\ Chap.~\bibinfo {chapter} {2}\BibitemShut {NoStop}%
\bibitem [{\citenamefont {McWeeny}(1956)}]{Mcweeny56}%
  \BibitemOpen
  \bibfield  {author} {\bibinfo {author} {\bibfnamefont {R.}~\bibnamefont
  {McWeeny}},\ }\href@noop {} {\bibfield  {journal} {\bibinfo  {journal} {Proc.
  R. Soc. A}\ }\textbf {\bibinfo {volume} {235}},\ \bibinfo {pages} {496}
  (\bibinfo {year} {1956})}\BibitemShut {NoStop}%
\bibitem [{\citenamefont {Fletcher}\ and\ \citenamefont
  {Reeves}(1964)}]{Fletcher64}%
  \BibitemOpen
  \bibfield  {author} {\bibinfo {author} {\bibfnamefont {R.}~\bibnamefont
  {Fletcher}}\ and\ \bibinfo {author} {\bibfnamefont {C.~M.}\ \bibnamefont
  {Reeves}},\ }\href@noop {} {\bibfield  {journal} {\bibinfo  {journal} {The
  Computer Journal}\ }\textbf {\bibinfo {volume} {7}},\ \bibinfo {pages} {149}
  (\bibinfo {year} {1964})}\BibitemShut {NoStop}%
\bibitem [{\citenamefont {Hillier}\ and\ \citenamefont
  {Saunders}(1970)}]{Hillier70}%
  \BibitemOpen
  \bibfield  {author} {\bibinfo {author} {\bibfnamefont {I.}~\bibnamefont
  {Hillier}}\ and\ \bibinfo {author} {\bibfnamefont {V.}~\bibnamefont
  {Saunders}},\ }\href@noop {} {\bibfield  {journal} {\bibinfo  {journal}
  {Proc. R. Soc. A}\ }\textbf {\bibinfo {volume} {320}},\ \bibinfo {pages}
  {161} (\bibinfo {year} {1970})}\BibitemShut {NoStop}%
\bibitem [{\citenamefont {Igawa}\ and\ \citenamefont
  {Fukutome}(1975)}]{Igawa75}%
  \BibitemOpen
  \bibfield  {author} {\bibinfo {author} {\bibfnamefont {A.}~\bibnamefont
  {Igawa}}\ and\ \bibinfo {author} {\bibfnamefont {H.}~\bibnamefont
  {Fukutome}},\ }\href {\doibase 10.1143/PTP.54.1266} {\bibfield  {journal}
  {\bibinfo  {journal} {Prog. Theor. Phys.}\ }\textbf {\bibinfo {volume}
  {54}},\ \bibinfo {pages} {1266} (\bibinfo {year} {1975})}\BibitemShut
  {NoStop}%
\bibitem [{\citenamefont {Seeger}\ and\ \citenamefont
  {Pople}(1976)}]{Seeger76}%
  \BibitemOpen
  \bibfield  {author} {\bibinfo {author} {\bibfnamefont {R.}~\bibnamefont
  {Seeger}}\ and\ \bibinfo {author} {\bibfnamefont {J.~A.}\ \bibnamefont
  {Pople}},\ }\href {\doibase 10.1063/1.432764} {\bibfield  {journal} {\bibinfo
   {journal} {J. Chem. Phys.}\ }\textbf {\bibinfo {volume} {65}},\ \bibinfo
  {pages} {265} (\bibinfo {year} {1976})}\BibitemShut {NoStop}%
\bibitem [{\citenamefont {Camp}\ and\ \citenamefont {King}(1981)}]{Camp81}%
  \BibitemOpen
  \bibfield  {author} {\bibinfo {author} {\bibfnamefont {R.~N.}\ \bibnamefont
  {Camp}}\ and\ \bibinfo {author} {\bibfnamefont {H.~F.}\ \bibnamefont
  {King}},\ }\href {\doibase 10.1063/1.441834} {\bibfield  {journal} {\bibinfo
  {journal} {J. Chem. Phys.}\ }\textbf {\bibinfo {volume} {75}},\ \bibinfo
  {pages} {268} (\bibinfo {year} {1981})}\BibitemShut {NoStop}%
\bibitem [{\citenamefont {Stanton}(1981)}]{Stanton81}%
  \BibitemOpen
  \bibfield  {author} {\bibinfo {author} {\bibfnamefont {R.~E.}\ \bibnamefont
  {Stanton}},\ }\href {\doibase 10.1063/1.442451} {\bibfield  {journal}
  {\bibinfo  {journal} {J. Chem. Phys.}\ }\textbf {\bibinfo {volume} {75}},\
  \bibinfo {pages} {3426} (\bibinfo {year} {1981})}\BibitemShut {NoStop}%
\bibitem [{\citenamefont {Bacskay}(1981)}]{Bacskay81}%
  \BibitemOpen
  \bibfield  {author} {\bibinfo {author} {\bibfnamefont {G.~B.}\ \bibnamefont
  {Bacskay}},\ }\href {\doibase 10.1016/0301-0104(81)85156-7} {\bibfield
  {journal} {\bibinfo  {journal} {Chem. Phys.}\ }\textbf {\bibinfo {volume}
  {61}},\ \bibinfo {pages} {385 } (\bibinfo {year} {1981})}\BibitemShut
  {NoStop}%
\bibitem [{\citenamefont {Cances}\ and\ \citenamefont {Le~Bris}(2000)}]{CL00}%
  \BibitemOpen
  \bibfield  {author} {\bibinfo {author} {\bibfnamefont {E.}~\bibnamefont
  {Cances}}\ and\ \bibinfo {author} {\bibfnamefont {C.}~\bibnamefont
  {Le~Bris}},\ }\href@noop {} {\bibfield  {journal} {\bibinfo  {journal} {Int.
  J. Quantum Chem.}\ }\textbf {\bibinfo {volume} {79}},\ \bibinfo {pages} {82}
  (\bibinfo {year} {2000})}\BibitemShut {NoStop}%
\bibitem [{\citenamefont {Kudin}, \citenamefont {Scuseria},\ and\ \citenamefont
  {Cances}(2002)}]{KSC02}%
  \BibitemOpen
  \bibfield  {author} {\bibinfo {author} {\bibfnamefont {K.}~\bibnamefont
  {Kudin}}, \bibinfo {author} {\bibfnamefont {G.}~\bibnamefont {Scuseria}}, \
  and\ \bibinfo {author} {\bibfnamefont {E.}~\bibnamefont {Cances}},\ }\href
  {\doibase 10.1063/1.1470195} {\bibfield  {journal} {\bibinfo  {journal} {J.
  Chem. Phys.}\ }\textbf {\bibinfo {volume} {116}},\ \bibinfo {pages} {8255}
  (\bibinfo {year} {2002})}\BibitemShut {NoStop}%
\bibitem [{\citenamefont {Thogersen}\ \emph {et~al.}(2004)\citenamefont
  {Thogersen}, \citenamefont {Olsen}, \citenamefont {Yeager}, \citenamefont
  {Jorgensen}, \citenamefont {Salek},\ and\ \citenamefont {Helgaker}}]{T04}%
  \BibitemOpen
  \bibfield  {author} {\bibinfo {author} {\bibfnamefont {L.}~\bibnamefont
  {Thogersen}}, \bibinfo {author} {\bibfnamefont {J.}~\bibnamefont {Olsen}},
  \bibinfo {author} {\bibfnamefont {D.}~\bibnamefont {Yeager}}, \bibinfo
  {author} {\bibfnamefont {P.}~\bibnamefont {Jorgensen}}, \bibinfo {author}
  {\bibfnamefont {P.}~\bibnamefont {Salek}}, \ and\ \bibinfo {author}
  {\bibfnamefont {T.}~\bibnamefont {Helgaker}},\ }\href {\doibase
  10.1063/1.1755673} {\bibfield  {journal} {\bibinfo  {journal} {J. Chem.
  Phys.}\ }\textbf {\bibinfo {volume} {121}},\ \bibinfo {pages} {16} (\bibinfo
  {year} {2004})}\BibitemShut {NoStop}%
\bibitem [{\citenamefont {Francisco}, \citenamefont {Martinez},\ and\
  \citenamefont {Martinez}(2004)}]{FMM04}%
  \BibitemOpen
  \bibfield  {author} {\bibinfo {author} {\bibfnamefont {J.~B.}\ \bibnamefont
  {Francisco}}, \bibinfo {author} {\bibfnamefont {J.~M.}\ \bibnamefont
  {Martinez}}, \ and\ \bibinfo {author} {\bibfnamefont {L.}~\bibnamefont
  {Martinez}},\ }\href {\doibase 10.1063/1.1814935} {\bibfield  {journal}
  {\bibinfo  {journal} {J. Chem. Phys.}\ }\textbf {\bibinfo {volume} {121}},\
  \bibinfo {pages} {10863} (\bibinfo {year} {2004})}\BibitemShut {NoStop}%
\bibitem [{\citenamefont {Francisco}, \citenamefont {Martinez},\ and\
  \citenamefont {Martinez}(2006)}]{FMM06}%
  \BibitemOpen
  \bibfield  {author} {\bibinfo {author} {\bibfnamefont {J.}~\bibnamefont
  {Francisco}}, \bibinfo {author} {\bibfnamefont {J.}~\bibnamefont {Martinez}},
  \ and\ \bibinfo {author} {\bibfnamefont {L.}~\bibnamefont {Martinez}},\
  }\href {http://dx.doi.org/10.1007/s10910-006-9058-0} {\bibfield  {journal}
  {\bibinfo  {journal} {J. Math. Chem.}\ }\textbf {\bibinfo {volume} {40}},\
  \bibinfo {pages} {349} (\bibinfo {year} {2006})}\BibitemShut {NoStop}%
\bibitem [{\citenamefont {Mazziotti}(2007)}]{RDM07}%
  \BibitemOpen
  \bibinfo {editor} {\bibfnamefont {D.~A.}\ \bibnamefont {Mazziotti}},\ ed.,\
  \href@noop {} {\emph {\bibinfo {title} {Reduced-Density-Matrix Mechanics:
  With Application to Many-electron Atoms and Molecules}}},\ \bibinfo {series}
  {Advances in Chemical Physics}, Vol.\ \bibinfo {volume} {134}\ (\bibinfo
  {publisher} {Wiley},\ \bibinfo {address} {New York},\ \bibinfo {year}
  {2007})\BibitemShut {NoStop}%
\bibitem [{\citenamefont {Coleman}\ and\ \citenamefont {Yukalov}(2000)}]{CB}%
  \BibitemOpen
  \bibfield  {author} {\bibinfo {author} {\bibfnamefont {A.~J.}\ \bibnamefont
  {Coleman}}\ and\ \bibinfo {author} {\bibfnamefont {V.~I.}\ \bibnamefont
  {Yukalov}},\ }\href@noop {} {\emph {\bibinfo {title} {Reduced Density
  Matrices: Coulson's Challenge}}}\ (\bibinfo  {publisher} {Springer-Verlag},\
  \bibinfo {address} {New York},\ \bibinfo {year} {2000})\BibitemShut {NoStop}%
\bibitem [{\citenamefont {Mazziotti}(2012{\natexlab{a}})}]{Nrep2}%
  \BibitemOpen
  \bibfield  {author} {\bibinfo {author} {\bibfnamefont {D.~A.}\ \bibnamefont
  {Mazziotti}},\ }\href {\doibase 10.1103/PhysRevLett.108.263002} {\bibfield
  {journal} {\bibinfo  {journal} {Phys. Rev. Lett.}\ }\textbf {\bibinfo
  {volume} {108}},\ \bibinfo {pages} {263002} (\bibinfo {year}
  {2012}{\natexlab{a}})}\BibitemShut {NoStop}%
\bibitem [{\citenamefont {Erdahl}(1979)}]{E79}%
  \BibitemOpen
  \bibfield  {author} {\bibinfo {author} {\bibfnamefont {R.~M.}\ \bibnamefont
  {Erdahl}},\ }\href@noop {} {\bibfield  {journal} {\bibinfo  {journal}
  {Reports Math. Phys.}\ }\textbf {\bibinfo {volume} {15}},\ \bibinfo {pages}
  {147} (\bibinfo {year} {1979})}\BibitemShut {NoStop}%
\bibitem [{\citenamefont {Mazziotti}\ and\ \citenamefont
  {Erdahl}(2001)}]{ME01}%
  \BibitemOpen
  \bibfield  {author} {\bibinfo {author} {\bibfnamefont {D.~A.}\ \bibnamefont
  {Mazziotti}}\ and\ \bibinfo {author} {\bibfnamefont {R.~M.}\ \bibnamefont
  {Erdahl}},\ }\href@noop {} {\bibfield  {journal} {\bibinfo  {journal} {Phys.
  Rev. A}\ }\textbf {\bibinfo {volume} {63}},\ \bibinfo {pages} {042113}
  (\bibinfo {year} {2001})}\BibitemShut {NoStop}%
\bibitem [{\citenamefont {Nakata}\ \emph {et~al.}(2001)\citenamefont {Nakata},
  \citenamefont {Nakatsuji}, \citenamefont {Ehara}, \citenamefont {Fukuda},
  \citenamefont {Nakata},\ and\ \citenamefont {Fujisawa}}]{N01}%
  \BibitemOpen
  \bibfield  {author} {\bibinfo {author} {\bibfnamefont {M.}~\bibnamefont
  {Nakata}}, \bibinfo {author} {\bibfnamefont {H.}~\bibnamefont {Nakatsuji}},
  \bibinfo {author} {\bibfnamefont {M.}~\bibnamefont {Ehara}}, \bibinfo
  {author} {\bibfnamefont {M.}~\bibnamefont {Fukuda}}, \bibinfo {author}
  {\bibfnamefont {K.}~\bibnamefont {Nakata}}, \ and\ \bibinfo {author}
  {\bibfnamefont {K.}~\bibnamefont {Fujisawa}},\ }\href@noop {} {\bibfield
  {journal} {\bibinfo  {journal} {J. Chem. Phys.}\ }\textbf {\bibinfo {volume}
  {114}},\ \bibinfo {pages} {8282} (\bibinfo {year} {2001})}\BibitemShut
  {NoStop}%
\bibitem [{\citenamefont {Mazziotti}(2002)}]{M02}%
  \BibitemOpen
  \bibfield  {author} {\bibinfo {author} {\bibfnamefont {D.~A.}\ \bibnamefont
  {Mazziotti}},\ }\href@noop {} {\bibfield  {journal} {\bibinfo  {journal}
  {Phys. Rev. A}\ }\textbf {\bibinfo {volume} {65}},\ \bibinfo {pages} {062511}
  (\bibinfo {year} {2002})}\BibitemShut {NoStop}%
\bibitem [{\citenamefont {Mazziotti}(2012{\natexlab{b}})}]{M12}%
  \BibitemOpen
  \bibfield  {author} {\bibinfo {author} {\bibfnamefont {D.~A.}\ \bibnamefont
  {Mazziotti}},\ }\href@noop {} {\bibfield  {journal} {\bibinfo  {journal}
  {Chem. Rev.}\ }\textbf {\bibinfo {volume} {12}},\ \bibinfo {pages} {244}
  (\bibinfo {year} {2012}{\natexlab{b}})}\BibitemShut {NoStop}%
\bibitem [{\citenamefont {Garrod}, \citenamefont {Mihailovi{\'c}},\ and\
  \citenamefont {Rosina}(1975)}]{G75}%
  \BibitemOpen
  \bibfield  {author} {\bibinfo {author} {\bibfnamefont {C.}~\bibnamefont
  {Garrod}}, \bibinfo {author} {\bibfnamefont {V.}~\bibnamefont
  {Mihailovi{\'c}}}, \ and\ \bibinfo {author} {\bibfnamefont {M.}~\bibnamefont
  {Rosina}},\ }\href@noop {} {\bibfield  {journal} {\bibinfo  {journal} {J.
  Math. Phys.}\ }\textbf {\bibinfo {volume} {10}},\ \bibinfo {pages} {1855}
  (\bibinfo {year} {1975})}\BibitemShut {NoStop}%
\bibitem [{\citenamefont {Zhao}\ \emph {et~al.}(2004)\citenamefont {Zhao},
  \citenamefont {Braams}, \citenamefont {Fukuda}, \citenamefont {Overton},\
  and\ \citenamefont {Percus}}]{Z04}%
  \BibitemOpen
  \bibfield  {author} {\bibinfo {author} {\bibfnamefont {Z.}~\bibnamefont
  {Zhao}}, \bibinfo {author} {\bibfnamefont {B.~J.}\ \bibnamefont {Braams}},
  \bibinfo {author} {\bibfnamefont {H.}~\bibnamefont {Fukuda}}, \bibinfo
  {author} {\bibfnamefont {M.~L.}\ \bibnamefont {Overton}}, \ and\ \bibinfo
  {author} {\bibfnamefont {J.~K.}\ \bibnamefont {Percus}},\ }\href@noop {}
  {\bibfield  {journal} {\bibinfo  {journal} {J. Chem. Phys.}\ }\textbf
  {\bibinfo {volume} {120}},\ \bibinfo {pages} {2095} (\bibinfo {year}
  {2004})}\BibitemShut {NoStop}%
\bibitem [{\citenamefont {Mazziotti}(2004)}]{M04}%
  \BibitemOpen
  \bibfield  {author} {\bibinfo {author} {\bibfnamefont {D.~A.}\ \bibnamefont
  {Mazziotti}},\ }\href {\doibase 10.1103/PhysRevLett.93.213001} {\bibfield
  {journal} {\bibinfo  {journal} {Phys. Rev. Lett.}\ }\textbf {\bibinfo
  {volume} {93}},\ \bibinfo {pages} {213001} (\bibinfo {year}
  {2004})}\BibitemShut {NoStop}%
\bibitem [{\citenamefont {Mazziotti}(2006)}]{M06}%
  \BibitemOpen
  \bibfield  {author} {\bibinfo {author} {\bibfnamefont {D.~A.}\ \bibnamefont
  {Mazziotti}},\ }\href@noop {} {\bibfield  {journal} {\bibinfo  {journal}
  {Phys. Rev. A}\ }\textbf {\bibinfo {volume} {74}},\ \bibinfo {pages} {032501}
  (\bibinfo {year} {2006})}\BibitemShut {NoStop}%
\bibitem [{\citenamefont {Canc\'es}, \citenamefont {Stoltz},\ and\
  \citenamefont {Lewin}(2006)}]{C06}%
  \BibitemOpen
  \bibfield  {author} {\bibinfo {author} {\bibfnamefont {E.}~\bibnamefont
  {Canc\'es}}, \bibinfo {author} {\bibfnamefont {G.}~\bibnamefont {Stoltz}}, \
  and\ \bibinfo {author} {\bibfnamefont {M.}~\bibnamefont {Lewin}},\
  }\href@noop {} {\bibfield  {journal} {\bibinfo  {journal} {J. Chem. Phys.}\
  }\textbf {\bibinfo {volume} {125}},\ \bibinfo {pages} {064101} (\bibinfo
  {year} {2006})}\BibitemShut {NoStop}%
\bibitem [{\citenamefont {Shenvi}\ and\ \citenamefont {Izmaylov}(2010)}]{S10}%
  \BibitemOpen
  \bibfield  {author} {\bibinfo {author} {\bibfnamefont {N.}~\bibnamefont
  {Shenvi}}\ and\ \bibinfo {author} {\bibfnamefont {A.~F.}\ \bibnamefont
  {Izmaylov}},\ }\href@noop {} {\bibfield  {journal} {\bibinfo  {journal}
  {Phys. Rev. Lett.}\ }\textbf {\bibinfo {volume} {105}},\ \bibinfo {pages}
  {213003} (\bibinfo {year} {2010})}\BibitemShut {NoStop}%
\bibitem [{\citenamefont {Verstichel}\ \emph {et~al.}(2012)\citenamefont
  {Verstichel}, \citenamefont {van Aggelen}, \citenamefont {Poelmans},\ and\
  \citenamefont {Neck}}]{B12}%
  \BibitemOpen
  \bibfield  {author} {\bibinfo {author} {\bibfnamefont {B.}~\bibnamefont
  {Verstichel}}, \bibinfo {author} {\bibfnamefont {H.}~\bibnamefont {van
  Aggelen}}, \bibinfo {author} {\bibfnamefont {W.}~\bibnamefont {Poelmans}}, \
  and\ \bibinfo {author} {\bibfnamefont {D.~V.}\ \bibnamefont {Neck}},\
  }\href@noop {} {\bibfield  {journal} {\bibinfo  {journal} {Phys. Rev. Lett.}\
  }\textbf {\bibinfo {volume} {108}},\ \bibinfo {pages} {213001} (\bibinfo
  {year} {2012})}\BibitemShut {NoStop}%
\bibitem [{\citenamefont {Mazziotti}(2011)}]{M11}%
  \BibitemOpen
  \bibfield  {author} {\bibinfo {author} {\bibfnamefont {D.~A.}\ \bibnamefont
  {Mazziotti}},\ }\href {\doibase 10.1103/PhysRevLett.106.083001} {\bibfield
  {journal} {\bibinfo  {journal} {Phys. Rev. Lett.}\ }\textbf {\bibinfo
  {volume} {106}},\ \bibinfo {pages} {083001} (\bibinfo {year}
  {2011})}\BibitemShut {NoStop}%
\bibitem [{\citenamefont {Pelzer}\ \emph {et~al.}(2011)\citenamefont {Pelzer},
  \citenamefont {Greenman}, \citenamefont {Gidofalvi},\ and\ \citenamefont
  {Mazziotti}}]{Kenley}%
  \BibitemOpen
  \bibfield  {author} {\bibinfo {author} {\bibfnamefont {K.}~\bibnamefont
  {Pelzer}}, \bibinfo {author} {\bibfnamefont {L.}~\bibnamefont {Greenman}},
  \bibinfo {author} {\bibfnamefont {G.}~\bibnamefont {Gidofalvi}}, \ and\
  \bibinfo {author} {\bibfnamefont {D.~A.}\ \bibnamefont {Mazziotti}},\ }\href
  {\doibase 10.1021/jp2017192} {\bibfield  {journal} {\bibinfo  {journal} {J.
  Phys. Chem. A}\ }\textbf {\bibinfo {volume} {115}},\ \bibinfo {pages} {5632}
  (\bibinfo {year} {2011})}\BibitemShut {NoStop}%
\bibitem [{\citenamefont {Greenman}\ and\ \citenamefont
  {Mazziotti}(2010)}]{firefly}%
  \BibitemOpen
  \bibfield  {author} {\bibinfo {author} {\bibfnamefont {L.}~\bibnamefont
  {Greenman}}\ and\ \bibinfo {author} {\bibfnamefont {D.~A.}\ \bibnamefont
  {Mazziotti}},\ }\href {\doibase http://dx.doi.org/10.1063/1.3501250}
  {\bibfield  {journal} {\bibinfo  {journal} {J. Chem. Phys.}\ }\textbf
  {\bibinfo {volume} {133}},\ \bibinfo {eid} {164110} (\bibinfo {year}
  {2010})}\BibitemShut {NoStop}%
\bibitem [{\citenamefont {Rothman}\ and\ \citenamefont
  {Mazziotti}(2008)}]{Adam08}%
  \BibitemOpen
  \bibfield  {author} {\bibinfo {author} {\bibfnamefont {A.~E.}\ \bibnamefont
  {Rothman}}\ and\ \bibinfo {author} {\bibfnamefont {D.~A.}\ \bibnamefont
  {Mazziotti}},\ }\href {\doibase 10.1103/PhysRevA.78.032510} {\bibfield
  {journal} {\bibinfo  {journal} {Phys. Rev. A}\ }\textbf {\bibinfo {volume}
  {78}},\ \bibinfo {pages} {032510} (\bibinfo {year} {2008})}\BibitemShut
  {NoStop}%
\bibitem [{\citenamefont {Gidofalvi}\ and\ \citenamefont
  {Mazziotti}(2006)}]{Greg06}%
  \BibitemOpen
  \bibfield  {author} {\bibinfo {author} {\bibfnamefont {G.}~\bibnamefont
  {Gidofalvi}}\ and\ \bibinfo {author} {\bibfnamefont {D.~A.}\ \bibnamefont
  {Mazziotti}},\ }\href {\doibase 10.1103/PhysRevA.74.012501} {\bibfield
  {journal} {\bibinfo  {journal} {Phys. Rev. A}\ }\textbf {\bibinfo {volume}
  {74}},\ \bibinfo {pages} {012501} (\bibinfo {year} {2006})}\BibitemShut
  {NoStop}%
\bibitem [{\citenamefont {Hammond}\ and\ \citenamefont
  {Mazziotti}(2006)}]{Jeff06}%
  \BibitemOpen
  \bibfield  {author} {\bibinfo {author} {\bibfnamefont {J.~R.}\ \bibnamefont
  {Hammond}}\ and\ \bibinfo {author} {\bibfnamefont {D.~A.}\ \bibnamefont
  {Mazziotti}},\ }\href {\doibase 10.1103/PhysRevA.73.062505} {\bibfield
  {journal} {\bibinfo  {journal} {Phys. Rev. A}\ }\textbf {\bibinfo {volume}
  {73}},\ \bibinfo {pages} {062505} (\bibinfo {year} {2006})}\BibitemShut
  {NoStop}%
\bibitem [{\citenamefont {Coleman}(1963)}]{Nrep}%
  \BibitemOpen
  \bibfield  {author} {\bibinfo {author} {\bibfnamefont {A.~J.}\ \bibnamefont
  {Coleman}},\ }\href {\doibase 10.1103/RevModPhys.35.668} {\bibfield
  {journal} {\bibinfo  {journal} {Rev. Mod. Phys.}\ }\textbf {\bibinfo {volume}
  {35}},\ \bibinfo {pages} {668} (\bibinfo {year} {1963})}\BibitemShut
  {NoStop}%
\bibitem [{\citenamefont {Harriman}(1978)}]{H78}%
  \BibitemOpen
  \bibfield  {author} {\bibinfo {author} {\bibfnamefont {J.~E.}\ \bibnamefont
  {Harriman}},\ }\href@noop {} {\bibfield  {journal} {\bibinfo  {journal}
  {Phys. Rev. A.}\ }\textbf {\bibinfo {volume} {17}},\ \bibinfo {pages} {1249}
  (\bibinfo {year} {1978})}\BibitemShut {NoStop}%
\bibitem [{\citenamefont {Lieb}(1981)}]{Lieb81}%
  \BibitemOpen
  \bibfield  {author} {\bibinfo {author} {\bibfnamefont {E.~H.}\ \bibnamefont
  {Lieb}},\ }\href {\doibase 10.1103/PhysRevLett.46.457} {\bibfield  {journal}
  {\bibinfo  {journal} {Phys. Rev. Lett.}\ }\textbf {\bibinfo {volume} {46}},\
  \bibinfo {pages} {457} (\bibinfo {year} {1981})}\BibitemShut {NoStop}%
\bibitem [{\citenamefont {Dattorro}(2013)}]{DB}%
  \BibitemOpen
  \bibfield  {author} {\bibinfo {author} {\bibfnamefont {J.}~\bibnamefont
  {Dattorro}},\ }\href {https://ccrma.stanford.edu/~dattorro/0976401304.pdf}
  {\emph {\bibinfo {title} {Convex Optimization \& Euclidean Distance
  Geometry}}}\ (\bibinfo  {publisher} {Meboo},\ \bibinfo {address} {Palo
  Alto},\ \bibinfo {year} {2013})\ Chap.~\bibinfo {chapter} {4}, pp.\ \bibinfo
  {pages} {308--333}\BibitemShut {NoStop}%
\bibitem [{\citenamefont {{Purvis III}}\ and\ \citenamefont
  {Bartlett}(1982)}]{PB82}%
  \BibitemOpen
  \bibfield  {author} {\bibinfo {author} {\bibfnamefont {G.~D.}\ \bibnamefont
  {{Purvis III}}}\ and\ \bibinfo {author} {\bibfnamefont {R.~J.}\ \bibnamefont
  {Bartlett}},\ }\href@noop {} {\bibfield  {journal} {\bibinfo  {journal} {J.
  Chem. Phys.}\ }\textbf {\bibinfo {volume} {76}},\ \bibinfo {pages} {1910}
  (\bibinfo {year} {1982})}\BibitemShut {NoStop}%
\bibitem [{\citenamefont {Bartlett}\ and\ \citenamefont
  {Musia\l{}}(2007)}]{BM07}%
  \BibitemOpen
  \bibfield  {author} {\bibinfo {author} {\bibfnamefont {R.~J.}\ \bibnamefont
  {Bartlett}}\ and\ \bibinfo {author} {\bibfnamefont {M.}~\bibnamefont
  {Musia\l{}}},\ }\href@noop {} {\bibfield  {journal} {\bibinfo  {journal}
  {Rev. Mod. Phys.}\ }\textbf {\bibinfo {volume} {79}},\ \bibinfo {pages} {291}
  (\bibinfo {year} {2007})}\BibitemShut {NoStop}%
\bibitem [{\citenamefont {Piecuch}\ \emph {et~al.}(2002)\citenamefont
  {Piecuch}, \citenamefont {Kucharski}, \citenamefont {Kowalski},\ and\
  \citenamefont {Musiał}}]{CCSD}%
  \BibitemOpen
  \bibfield  {author} {\bibinfo {author} {\bibfnamefont {P.}~\bibnamefont
  {Piecuch}}, \bibinfo {author} {\bibfnamefont {S.~A.}\ \bibnamefont
  {Kucharski}}, \bibinfo {author} {\bibfnamefont {K.}~\bibnamefont {Kowalski}},
  \ and\ \bibinfo {author} {\bibfnamefont {M.}~\bibnamefont {Musiał}},\ }\href
  {\doibase 10.1016/S0010-4655(02)00598-2} {\bibfield  {journal} {\bibinfo
  {journal} {Comp. Phys. Comm.}\ }\textbf {\bibinfo {volume} {149}},\ \bibinfo
  {pages} {71 } (\bibinfo {year} {2002})}\BibitemShut {NoStop}%
\bibitem [{\citenamefont {Pataki}(1998)}]{P98}%
  \BibitemOpen
  \bibfield  {author} {\bibinfo {author} {\bibfnamefont {G.}~\bibnamefont
  {Pataki}},\ }\href {\doibase 10.1287/moor.23.2.339} {\bibfield  {journal}
  {\bibinfo  {journal} {Math. Oper. Res.}\ }\textbf {\bibinfo {volume} {23}},\
  \bibinfo {pages} {339} (\bibinfo {year} {1998})}\BibitemShut {NoStop}%
\bibitem [{\citenamefont {Barvinok}(1995)}]{B95}%
  \BibitemOpen
  \bibfield  {author} {\bibinfo {author} {\bibfnamefont {A.}~\bibnamefont
  {Barvinok}},\ }\href {\doibase 10.1007/BF02574037} {\bibfield  {journal}
  {\bibinfo  {journal} {Discrete Comput. Geom.}\ }\textbf {\bibinfo {volume}
  {13}},\ \bibinfo {pages} {189} (\bibinfo {year} {1995})}\BibitemShut
  {NoStop}%
\bibitem [{\citenamefont {Gidofalvi}\ and\ \citenamefont
  {Mazziotti}(2005)}]{GM05}%
  \BibitemOpen
  \bibfield  {author} {\bibinfo {author} {\bibfnamefont {G.}~\bibnamefont
  {Gidofalvi}}\ and\ \bibinfo {author} {\bibfnamefont {D.~A.}\ \bibnamefont
  {Mazziotti}},\ }\href {\doibase 10.1063/1.1901565} {\bibfield  {journal}
  {\bibinfo  {journal} {J. Chem. Phys.}\ }\textbf {\bibinfo {volume} {122}},\
  \bibinfo {eid} {194104} (\bibinfo {year} {2005})}\BibitemShut {NoStop}%
\bibitem [{\citenamefont {Abrams}\ and\ \citenamefont {Sherrill}(2004)}]{AS04}%
  \BibitemOpen
  \bibfield  {author} {\bibinfo {author} {\bibfnamefont {M.~L.}\ \bibnamefont
  {Abrams}}\ and\ \bibinfo {author} {\bibfnamefont {C.~D.}\ \bibnamefont
  {Sherrill}},\ }\href {\doibase 10.1063/1.1804498} {\bibfield  {journal}
  {\bibinfo  {journal} {J. Chem. Phys.}\ }\textbf {\bibinfo {volume} {121}},\
  \bibinfo {pages} {9211} (\bibinfo {year} {2004})}\BibitemShut {NoStop}%
\bibitem [{\citenamefont {Goodgame}\ and\ \citenamefont
  {Goddard}(1985)}]{Good85}%
  \BibitemOpen
  \bibfield  {author} {\bibinfo {author} {\bibfnamefont {M.~M.}\ \bibnamefont
  {Goodgame}}\ and\ \bibinfo {author} {\bibfnamefont {W.~A.}\ \bibnamefont
  {Goddard}},\ }\href {\doibase 10.1103/PhysRevLett.54.661} {\bibfield
  {journal} {\bibinfo  {journal} {Phys. Rev. Lett.}\ }\textbf {\bibinfo
  {volume} {54}},\ \bibinfo {pages} {661} (\bibinfo {year} {1985})}\BibitemShut
  {NoStop}%
\bibitem [{\citenamefont {Roos}(2003)}]{Roos03}%
  \BibitemOpen
  \bibfield  {author} {\bibinfo {author} {\bibfnamefont {B.~O.}\ \bibnamefont
  {Roos}},\ }\href@noop {} {\bibfield  {journal} {\bibinfo  {journal}
  {Collection of Czechoslovak Chemical Communications}\ }\textbf {\bibinfo
  {volume} {68}},\ \bibinfo {pages} {265} (\bibinfo {year} {2003})}\BibitemShut
  {NoStop}%
\bibitem [{\citenamefont {Celani}\ \emph {et~al.}(2004)\citenamefont {Celani},
  \citenamefont {Stoll}, \citenamefont {Werner},\ and\ \citenamefont
  {Knowles}}]{Celani04}%
  \BibitemOpen
  \bibfield  {author} {\bibinfo {author} {\bibfnamefont {P.}~\bibnamefont
  {Celani}}, \bibinfo {author} {\bibfnamefont {H.}~\bibnamefont {Stoll}},
  \bibinfo {author} {\bibfnamefont {H.-J.}\ \bibnamefont {Werner}}, \ and\
  \bibinfo {author} {\bibfnamefont {P.}~\bibnamefont {Knowles}},\ }\href@noop
  {} {\bibfield  {journal} {\bibinfo  {journal} {Mol. Phys.}\ }\textbf
  {\bibinfo {volume} {102}},\ \bibinfo {pages} {2369} (\bibinfo {year}
  {2004})}\BibitemShut {NoStop}%
\bibitem [{\citenamefont {Muller}(2009)}]{Muller09}%
  \BibitemOpen
  \bibfield  {author} {\bibinfo {author} {\bibfnamefont {T.}~\bibnamefont
  {Muller}},\ }\href {\doibase 10.1021/jp905254u} {\bibfield  {journal}
  {\bibinfo  {journal} {J. Phys. Chem. A}\ }\textbf {\bibinfo {volume} {113}},\
  \bibinfo {pages} {12729} (\bibinfo {year} {2009})}\BibitemShut {NoStop}%
\bibitem [{\citenamefont {Andersson}\ \emph {et~al.}(1994)\citenamefont
  {Andersson}, \citenamefont {Roos}, \citenamefont {Malmqvist},\ and\
  \citenamefont {Widmark}}]{And94}%
  \BibitemOpen
  \bibfield  {author} {\bibinfo {author} {\bibfnamefont {K.}~\bibnamefont
  {Andersson}}, \bibinfo {author} {\bibfnamefont {B.}~\bibnamefont {Roos}},
  \bibinfo {author} {\bibfnamefont {P.-A.}\ \bibnamefont {Malmqvist}}, \ and\
  \bibinfo {author} {\bibfnamefont {P.-O.}\ \bibnamefont {Widmark}},\ }\href
  {\doibase 10.1016/0009-2614(94)01183-4} {\bibfield  {journal} {\bibinfo
  {journal} {Chem. Phys. Lett.}\ }\textbf {\bibinfo {volume} {230}},\ \bibinfo
  {pages} {391 } (\bibinfo {year} {1994})}\BibitemShut {NoStop}%
\bibitem [{\citenamefont {{Bauschlicher Jr.}}\ and\ \citenamefont
  {Partridge}(1994)}]{Bausch94}%
  \BibitemOpen
  \bibfield  {author} {\bibinfo {author} {\bibfnamefont {C.~W.}\ \bibnamefont
  {{Bauschlicher Jr.}}}\ and\ \bibinfo {author} {\bibfnamefont
  {H.}~\bibnamefont {Partridge}},\ }\href {\doibase
  10.1016/0009-2614(94)01243-1} {\bibfield  {journal} {\bibinfo  {journal}
  {Chem. Phys. Lett.}\ }\textbf {\bibinfo {volume} {231}},\ \bibinfo {pages}
  {277 } (\bibinfo {year} {1994})}\BibitemShut {NoStop}%
\bibitem [{\citenamefont {Roos}\ and\ \citenamefont
  {Andersson}(1995)}]{Roos95}%
  \BibitemOpen
  \bibfield  {author} {\bibinfo {author} {\bibfnamefont {B.~O.}\ \bibnamefont
  {Roos}}\ and\ \bibinfo {author} {\bibfnamefont {K.}~\bibnamefont
  {Andersson}},\ }\href {\doibase 10.1016/0009-2614(95)01010-7} {\bibfield
  {journal} {\bibinfo  {journal} {Chem. Phys. Lett.}\ }\textbf {\bibinfo
  {volume} {245}},\ \bibinfo {pages} {215 } (\bibinfo {year}
  {1995})}\BibitemShut {NoStop}%
\bibitem [{\citenamefont {Zgid}\ \emph {et~al.}(2009)\citenamefont {Zgid},
  \citenamefont {Ghosh}, \citenamefont {Neuscamman},\ and\ \citenamefont
  {Chan}}]{Zgid09}%
  \BibitemOpen
  \bibfield  {author} {\bibinfo {author} {\bibfnamefont {D.}~\bibnamefont
  {Zgid}}, \bibinfo {author} {\bibfnamefont {D.}~\bibnamefont {Ghosh}},
  \bibinfo {author} {\bibfnamefont {E.}~\bibnamefont {Neuscamman}}, \ and\
  \bibinfo {author} {\bibfnamefont {G.~K.-L.}\ \bibnamefont {Chan}},\ }\href
  {\doibase 10.1063/1.3132922} {\bibfield  {journal} {\bibinfo  {journal} {J.
  Chem. Phys.}\ }\textbf {\bibinfo {volume} {130}},\ \bibinfo {eid} {194107}
  (\bibinfo {year} {2009})}\BibitemShut {NoStop}%
\bibitem [{\citenamefont {Angeli}\ \emph {et~al.}(2006)\citenamefont {Angeli},
  \citenamefont {Bories}, \citenamefont {Cavallini},\ and\ \citenamefont
  {Cimiraglia}}]{Angeli06}%
  \BibitemOpen
  \bibfield  {author} {\bibinfo {author} {\bibfnamefont {C.}~\bibnamefont
  {Angeli}}, \bibinfo {author} {\bibfnamefont {B.}~\bibnamefont {Bories}},
  \bibinfo {author} {\bibfnamefont {A.}~\bibnamefont {Cavallini}}, \ and\
  \bibinfo {author} {\bibfnamefont {R.}~\bibnamefont {Cimiraglia}},\ }\href
  {\doibase 10.1063/1.2148946} {\bibfield  {journal} {\bibinfo  {journal} {J.
  Chem. Phys.}\ }\textbf {\bibinfo {volume} {124}},\ \bibinfo {eid} {054108}
  (\bibinfo {year} {2006})}\BibitemShut {NoStop}%
\bibitem [{\citenamefont {Schäfer}, \citenamefont {Huber},\ and\ \citenamefont
  {Ahlrichs}(1994)}]{TZV}%
  \BibitemOpen
  \bibfield  {author} {\bibinfo {author} {\bibfnamefont {A.}~\bibnamefont
  {Schäfer}}, \bibinfo {author} {\bibfnamefont {C.}~\bibnamefont {Huber}}, \
  and\ \bibinfo {author} {\bibfnamefont {R.}~\bibnamefont {Ahlrichs}},\
  }\href@noop {} {\bibfield  {journal} {\bibinfo  {journal} {J. Chem. Phys.}\
  }\textbf {\bibinfo {volume} {100}} (\bibinfo {year} {1994})}\BibitemShut
  {NoStop}%
\bibitem [{\citenamefont {Mazziotti}(2008)}]{M08}%
  \BibitemOpen
  \bibfield  {author} {\bibinfo {author} {\bibfnamefont {D.~A.}\ \bibnamefont
  {Mazziotti}},\ }\href@noop {} {\bibfield  {journal} {\bibinfo  {journal}
  {Phys. Rev. Lett.}\ }\textbf {\bibinfo {volume} {101}},\ \bibinfo {pages}
  {253002} (\bibinfo {year} {2008})}\BibitemShut {NoStop}%
\bibitem [{\citenamefont {Mazziotti}(2010)}]{M10}%
  \BibitemOpen
  \bibfield  {author} {\bibinfo {author} {\bibfnamefont {D.~A.}\ \bibnamefont
  {Mazziotti}},\ }\href@noop {} {\bibfield  {journal} {\bibinfo  {journal}
  {Phys. Rev. A.}\ }\textbf {\bibinfo {volume} {81}},\ \bibinfo {pages}
  {062515} (\bibinfo {year} {2010})}\BibitemShut {NoStop}%
\bibitem [{\citenamefont {Scuseria}\ and\ \citenamefont {{Schaefer
  III}}(1990)}]{Cr2-CCSD}%
  \BibitemOpen
  \bibfield  {author} {\bibinfo {author} {\bibfnamefont {G.~E.}\ \bibnamefont
  {Scuseria}}\ and\ \bibinfo {author} {\bibfnamefont {H.~F.}\ \bibnamefont
  {{Schaefer III}}},\ }\href {\doibase 10.1016/S0009-2614(90)87186-U}
  {\bibfield  {journal} {\bibinfo  {journal} {Chem. Phys. Lett.}\ }\textbf
  {\bibinfo {volume} {174}},\ \bibinfo {pages} {501 } (\bibinfo {year}
  {1990})}\BibitemShut {NoStop}%
\bibitem [{\citenamefont {{Combi}}(1980)}]{C80}%
  \BibitemOpen
  \bibfield  {author} {\bibinfo {author} {\bibfnamefont {M.~R.}\ \bibnamefont
  {{Combi}}},\ }\href {\doibase 10.1086/158394} {\bibfield  {journal} {\bibinfo
   {journal} {Astrophys. J.}\ }\textbf {\bibinfo {volume} {241}},\ \bibinfo
  {pages} {830} (\bibinfo {year} {1980})}\BibitemShut {NoStop}%
\bibitem [{\citenamefont {Knowles}\ \emph {et~al.}(1988)\citenamefont
  {Knowles}, \citenamefont {Werner}, \citenamefont {Hay},\ and\ \citenamefont
  {Cartwright}}]{KWHC88}%
  \BibitemOpen
  \bibfield  {author} {\bibinfo {author} {\bibfnamefont {P.~J.}\ \bibnamefont
  {Knowles}}, \bibinfo {author} {\bibfnamefont {H.-J.}\ \bibnamefont {Werner}},
  \bibinfo {author} {\bibfnamefont {P.~J.}\ \bibnamefont {Hay}}, \ and\
  \bibinfo {author} {\bibfnamefont {D.~C.}\ \bibnamefont {Cartwright}},\ }\href
  {\doibase 10.1063/1.455264} {\bibfield  {journal} {\bibinfo  {journal} {J.
  Chem. Phys.}\ }\textbf {\bibinfo {volume} {89}},\ \bibinfo {pages} {7334}
  (\bibinfo {year} {1988})}\BibitemShut {NoStop}%
\bibitem [{\citenamefont {Mele}\ and\ \citenamefont {Okabe}(1969)}]{MO69}%
  \BibitemOpen
  \bibfield  {author} {\bibinfo {author} {\bibfnamefont {A.}~\bibnamefont
  {Mele}}\ and\ \bibinfo {author} {\bibfnamefont {H.}~\bibnamefont {Okabe}},\
  }\href {\doibase 10.1063/1.1671870} {\bibfield  {journal} {\bibinfo
  {journal} {J. Chem. Phys.}\ }\textbf {\bibinfo {volume} {51}},\ \bibinfo
  {pages} {4798} (\bibinfo {year} {1969})}\BibitemShut {NoStop}%
\bibitem [{\citenamefont {West}\ and\ \citenamefont {Berry}(1974)}]{WB74}%
  \BibitemOpen
  \bibfield  {author} {\bibinfo {author} {\bibfnamefont {G.~A.}\ \bibnamefont
  {West}}\ and\ \bibinfo {author} {\bibfnamefont {M.~J.}\ \bibnamefont
  {Berry}},\ }\href {\doibase 10.1063/1.1681793} {\bibfield  {journal}
  {\bibinfo  {journal} {J. Chem. Phys.}\ }\textbf {\bibinfo {volume} {61}},\
  \bibinfo {pages} {4700} (\bibinfo {year} {1974})}\BibitemShut {NoStop}%
\bibitem [{\citenamefont {Dunning{, Jr.}}(1989)}]{D89}%
  \BibitemOpen
  \bibfield  {author} {\bibinfo {author} {\bibfnamefont {T.~H.}\ \bibnamefont
  {Dunning{, Jr.}}},\ }\href@noop {} {\bibfield  {journal} {\bibinfo  {journal}
  {J. Chem. Phys.}\ }\textbf {\bibinfo {volume} {90}},\ \bibinfo {pages} {1007}
  (\bibinfo {year} {1989})}\BibitemShut {NoStop}%
\bibitem [{\citenamefont {Thøgersen}\ and\ \citenamefont
  {Olsen}(2004)}]{TO04}%
  \BibitemOpen
  \bibfield  {author} {\bibinfo {author} {\bibfnamefont {L.}~\bibnamefont
  {Thøgersen}}\ and\ \bibinfo {author} {\bibfnamefont {J.}~\bibnamefont
  {Olsen}},\ }\href {\doibase http://dx.doi.org/10.1016/j.cplett.2004.06.001}
  {\bibfield  {journal} {\bibinfo  {journal} {Chem. Phys. Lett.}\ }\textbf
  {\bibinfo {volume} {393}},\ \bibinfo {pages} {36 } (\bibinfo {year}
  {2004})}\BibitemShut {NoStop}%
\bibitem [{\citenamefont {Wilkinson}\ and\ \citenamefont
  {J.~Whitaker}(2010)}]{WW10}%
  \BibitemOpen
  \bibfield  {author} {\bibinfo {author} {\bibfnamefont {I.}~\bibnamefont
  {Wilkinson}}\ and\ \bibinfo {author} {\bibfnamefont {B.}~\bibnamefont
  {J.~Whitaker}},\ }\href {\doibase 10.1039/B924653N} {\bibfield  {journal}
  {\bibinfo  {journal} {Annu. Rep. Prog. Chem.{,} Sect. C: Phys. Chem.}\
  }\textbf {\bibinfo {volume} {106}},\ \bibinfo {pages} {274} (\bibinfo {year}
  {2010})}\BibitemShut {NoStop}%
\bibitem [{\citenamefont {Worner}\ \emph {et~al.}(2011)\citenamefont {Worner},
  \citenamefont {Bertrand}, \citenamefont {Fabre}, \citenamefont {Higuet},
  \citenamefont {Ruf}, \citenamefont {Dubrouil}, \citenamefont {Patchkovskii},
  \citenamefont {Spanner}, \citenamefont {Mairesse}, \citenamefont {Blanchet},
  \citenamefont {Mével}, \citenamefont {Constant}, \citenamefont {Corkum},\
  and\ \citenamefont {Villeneuve}}]{Worner11}%
  \BibitemOpen
  \bibfield  {author} {\bibinfo {author} {\bibfnamefont {H.~J.}\ \bibnamefont
  {Worner}}, \bibinfo {author} {\bibfnamefont {J.~B.}\ \bibnamefont
  {Bertrand}}, \bibinfo {author} {\bibfnamefont {B.}~\bibnamefont {Fabre}},
  \bibinfo {author} {\bibfnamefont {J.}~\bibnamefont {Higuet}}, \bibinfo
  {author} {\bibfnamefont {H.}~\bibnamefont {Ruf}}, \bibinfo {author}
  {\bibfnamefont {A.}~\bibnamefont {Dubrouil}}, \bibinfo {author}
  {\bibfnamefont {S.}~\bibnamefont {Patchkovskii}}, \bibinfo {author}
  {\bibfnamefont {M.}~\bibnamefont {Spanner}}, \bibinfo {author} {\bibfnamefont
  {Y.}~\bibnamefont {Mairesse}}, \bibinfo {author} {\bibfnamefont
  {V.}~\bibnamefont {Blanchet}}, \bibinfo {author} {\bibfnamefont
  {E.}~\bibnamefont {Mével}}, \bibinfo {author} {\bibfnamefont
  {E.}~\bibnamefont {Constant}}, \bibinfo {author} {\bibfnamefont {P.~B.}\
  \bibnamefont {Corkum}}, \ and\ \bibinfo {author} {\bibfnamefont {D.~M.}\
  \bibnamefont {Villeneuve}},\ }\href {\doibase 10.1126/science.1208664}
  {\bibfield  {journal} {\bibinfo  {journal} {Science}\ }\textbf {\bibinfo
  {volume} {334}},\ \bibinfo {pages} {208} (\bibinfo {year}
  {2011})}\BibitemShut {NoStop}%
\bibitem [{\citenamefont {Kraus}\ \emph {et~al.}(2012)\citenamefont {Kraus},
  \citenamefont {Arasaki}, \citenamefont {Bertrand}, \citenamefont
  {Patchkovskii}, \citenamefont {Corkum}, \citenamefont {Villeneuve},
  \citenamefont {Takatsuka},\ and\ \citenamefont {W\"orner}}]{Kraus12}%
  \BibitemOpen
  \bibfield  {author} {\bibinfo {author} {\bibfnamefont {P.~M.}\ \bibnamefont
  {Kraus}}, \bibinfo {author} {\bibfnamefont {Y.}~\bibnamefont {Arasaki}},
  \bibinfo {author} {\bibfnamefont {J.~B.}\ \bibnamefont {Bertrand}}, \bibinfo
  {author} {\bibfnamefont {S.}~\bibnamefont {Patchkovskii}}, \bibinfo {author}
  {\bibfnamefont {P.~B.}\ \bibnamefont {Corkum}}, \bibinfo {author}
  {\bibfnamefont {D.~M.}\ \bibnamefont {Villeneuve}}, \bibinfo {author}
  {\bibfnamefont {K.}~\bibnamefont {Takatsuka}}, \ and\ \bibinfo {author}
  {\bibfnamefont {H.~J.}\ \bibnamefont {W\"orner}},\ }\href {\doibase
  10.1103/PhysRevA.85.043409} {\bibfield  {journal} {\bibinfo  {journal} {Phys.
  Rev. A}\ }\textbf {\bibinfo {volume} {85}},\ \bibinfo {pages} {043409}
  (\bibinfo {year} {2012})}\BibitemShut {NoStop}%
\bibitem [{\citenamefont {\v{C}\'{\i}\v{z}ek}\ and\ \citenamefont
  {Paldus}(1967)}]{CP67}%
  \BibitemOpen
  \bibfield  {author} {\bibinfo {author} {\bibfnamefont {J.}~\bibnamefont
  {\v{C}\'{\i}\v{z}ek}}\ and\ \bibinfo {author} {\bibfnamefont
  {J.}~\bibnamefont {Paldus}},\ }\href {\doibase 10.1063/1.1701562} {\bibfield
  {journal} {\bibinfo  {journal} {J. Chem. Phys.}\ }\textbf {\bibinfo {volume}
  {47}},\ \bibinfo {pages} {3976} (\bibinfo {year} {1967})}\BibitemShut
  {NoStop}%
\bibitem [{\citenamefont {Seeger}\ and\ \citenamefont {Pople}(1977)}]{SP77}%
  \BibitemOpen
  \bibfield  {author} {\bibinfo {author} {\bibfnamefont {R.}~\bibnamefont
  {Seeger}}\ and\ \bibinfo {author} {\bibfnamefont {J.~A.}\ \bibnamefont
  {Pople}},\ }\href {\doibase 10.1063/1.434318} {\bibfield  {journal} {\bibinfo
   {journal} {J. Chem. Phys.}\ }\textbf {\bibinfo {volume} {66}},\ \bibinfo
  {pages} {3045} (\bibinfo {year} {1977})}\BibitemShut {NoStop}%
\bibitem [{\citenamefont {Jiminez-Hoyos}, \citenamefont {Rodriguez-Guzman},\
  and\ \citenamefont {Scuseria}(2013)}]{Scu13}%
  \BibitemOpen
  \bibfield  {author} {\bibinfo {author} {\bibfnamefont {C.~A.}\ \bibnamefont
  {Jiminez-Hoyos}}, \bibinfo {author} {\bibfnamefont {R.}~\bibnamefont
  {Rodriguez-Guzman}}, \ and\ \bibinfo {author} {\bibfnamefont {G.~E.}\
  \bibnamefont {Scuseria}},\ }\href {\doibase
  http://dx.doi.org/10.1063/1.4832476} {\bibfield  {journal} {\bibinfo
  {journal} {J. Chem. Phys.}\ }\textbf {\bibinfo {volume} {139}},\ \bibinfo
  {eid} {204102} (\bibinfo {year} {2013})}\BibitemShut {NoStop}%
\bibitem [{\citenamefont {Scuseria}\ \emph {et~al.}(2011)\citenamefont
  {Scuseria}, \citenamefont {Jimenez-Hoyos}, \citenamefont {Henderson},
  \citenamefont {Samanta},\ and\ \citenamefont {Ellis}}]{SJH11}%
  \BibitemOpen
  \bibfield  {author} {\bibinfo {author} {\bibfnamefont {G.~E.}\ \bibnamefont
  {Scuseria}}, \bibinfo {author} {\bibfnamefont {C.~A.}\ \bibnamefont
  {Jimenez-Hoyos}}, \bibinfo {author} {\bibfnamefont {T.~M.}\ \bibnamefont
  {Henderson}}, \bibinfo {author} {\bibfnamefont {K.}~\bibnamefont {Samanta}},
  \ and\ \bibinfo {author} {\bibfnamefont {J.~K.}\ \bibnamefont {Ellis}},\
  }\href@noop {} {\bibfield  {journal} {\bibinfo  {journal} {J. Chem. Phys.}\
  }\textbf {\bibinfo {volume} {135}},\ \bibinfo {pages} {124108} (\bibinfo
  {year} {2011})}\BibitemShut {NoStop}%
\bibitem [{\citenamefont {Mazziotti}(2000)}]{M00}%
  \BibitemOpen
  \bibfield  {author} {\bibinfo {author} {\bibfnamefont {D.~A.}\ \bibnamefont
  {Mazziotti}},\ }\href@noop {} {\bibfield  {journal} {\bibinfo  {journal} {J.
  Chem. Phys.}\ }\textbf {\bibinfo {volume} {112}} (\bibinfo {year}
  {2000})}\BibitemShut {NoStop}%
\bibitem [{\citenamefont {Mazziotti}(2001)}]{M01b}%
  \BibitemOpen
  \bibfield  {author} {\bibinfo {author} {\bibfnamefont {D.~A.}\ \bibnamefont
  {Mazziotti}},\ }\href {\doibase
  http://dx.doi.org/10.1016/S0009-2614(01)00251-2} {\bibfield  {journal}
  {\bibinfo  {journal} {Chem. Phys. Lett.}\ }\textbf {\bibinfo {volume}
  {338}},\ \bibinfo {pages} {323 } (\bibinfo {year} {2001})}\BibitemShut
  {NoStop}%
\bibitem [{\citenamefont {Broer}\ and\ \citenamefont
  {Nieuwpoort}(1988)}]{NOCI}%
  \BibitemOpen
  \bibfield  {author} {\bibinfo {author} {\bibfnamefont {R.}~\bibnamefont
  {Broer}}\ and\ \bibinfo {author} {\bibfnamefont {W.}~\bibnamefont
  {Nieuwpoort}},\ }\href {\doibase 10.1007/BF00527744} {\bibfield  {journal}
  {\bibinfo  {journal} {Theor. chim. acta}\ }\textbf {\bibinfo {volume} {73}},\
  \bibinfo {pages} {405} (\bibinfo {year} {1988})}\BibitemShut {NoStop}%
\bibitem [{\citenamefont {Hollauer}\ and\ \citenamefont
  {Nascimento}(1993)}]{NOCI2}%
  \BibitemOpen
  \bibfield  {author} {\bibinfo {author} {\bibfnamefont {E.}~\bibnamefont
  {Hollauer}}\ and\ \bibinfo {author} {\bibfnamefont {M.~A.~C.}\ \bibnamefont
  {Nascimento}},\ }\href@noop {} {\bibfield  {journal} {\bibinfo  {journal} {J.
  Chem. Phys.}\ }\textbf {\bibinfo {volume} {99}} (\bibinfo {year}
  {1993})}\BibitemShut {NoStop}%
\end{thebibliography}

%

\end{document}